\documentclass[aps,prd,showpacs,superscriptaddress,preprint,raggedfooter]{revtex4-1}

\usepackage[dvips]{graphicx}
\usepackage{bm}        
\usepackage{amssymb}   
\usepackage[utf8]{inputenc}
\usepackage{mathrsfs}
\usepackage{float}
\usepackage{amsmath}
\usepackage{graphicx,subfigure}
\usepackage{color}
\usepackage{multirow}
\usepackage[breaklinks=true,colorlinks=true]{hyperref}
\hypersetup{colorlinks=true,citecolor=blue,linkcolor=blue,urlcolor=blue}

\hypersetup{colorlinks=true,citecolor=blue,linkcolor=blue,urlcolor=blue}

\usepackage[english]{babel}
\usepackage{amsmath}
\begin{document}

\title{
Kink dynamics in a high-order field model
}

\author{Aliakbar Moradi Marjaneh}
\email{moradimarjaneh@iau.ac.ir}
\affiliation{Department of Physics, Qu.C., Islamic Azad University, Quchan, Iran}

\author{Vakhid A. Gani}
\email{vagani@mephi.ru}
\affiliation{
National Research Nuclear University MEPhI\\ (Moscow Engineering Physics Institute), Moscow 115409, Russia}
\affiliation{National Research Centre ``Kurchatov Institute'', Moscow 123182, Russia}

\author{Azam Ghaani}
\email{az.ghaani@alumni.um.ac.ir}
\affiliation{Department of Physics, Faculty of Sciences, Ferdowsi University of Mashhad, Mashhad, Iran}

\author{Kurosh Javidan}
\email{javidan@um.ac.ir}
\affiliation{Department of Physics,  Faculty of Sciences, Ferdowsi University of Mashhad, Mashhad, Iran}

\author{Alexander A. Malnev}
\email{malnev.aa@phystech.edu}
\affiliation{Moscow Institute of Physics and Technology,
Dolgoprudny, Moscow Region 141700, Russia}

\author{Oleg V. Nagornov}
\email{ovnagornov@mephi.ru}
\affiliation{
National Research Nuclear University MEPhI\\ (Moscow Engineering Physics Institute), Moscow 115409, Russia}


\begin{abstract}

We study various properties of topological solitons (kinks) of a field-theoretic model with a polynomial potential of the twelfth degree. This model is remarkable in that it has several topological sectors, in which kinks have different masses. We obtain asymptotic estimates for the kink-antikink and antikink-kink interaction forces. We also study numerically kink-antikink and antikink-kink collisions and observe a number of interesting phenomena: annihilation of a kink-antikink pair in one topological sector and the production in its place of a pair in another sector; resonance phenomena --- escape windows, despite the absence of vibrational modes in the kink excitation spectra.

\end{abstract}

\maketitle

\section{Introduction}\label{sec:introduction}
\label{sec:Introduction}

Field-theoretic models with non-linear self-interaction of a real scalar field evolving in $(1+1)$-dimensional space-time and, in particular, their topological solitons (kink solutions, or {\it kinks}) are of great importance for modern physics. Kinks and kink-like field configurations arise in many physical contexts. A very common example is a flat section of a cosmological domain wall separating space regions with different vacuums --- in the direction perpendicular to the wall, this is a kink-type field configuration \cite{Manton.book.2004,Vachaspati.book.2006,Shnir.book.2018}. In condensed matter, a wall separating two magnetic domains can be deformed in such a way that its shape locally reproduces the kink profile. Such a deformation moves along the wall, like a kink of a continuous field-theoretic model \cite{Buijnsters.PRL.2014}. Another striking example of a kink is the deformation of a long narrow sample --- a graphene nanoribbon \cite{Yamaletdinov.PRB.2017,Yamaletdinov.Carbon.2019,Nguyen.PRB.2021,Nguyen.PRApplied.2022}.

Kink solutions arise in models that describe sequences of phase transitions. Details as well as some literature review can be found in \cite{Khare.PRE.2014}, see also chapter 12 in \cite{Kevrekidis.book.2019}. The $\varphi^{12}$ model, which this paper is devoted to, was used to describe the phenomenology of phase transitions in highly piezoelectric perovskite materials \cite{Vanderbilt.PRB.2001,Sergienko.PRB.2002}.

The best-known models seem to be the integrable sine-Gordon model \cite{Cuevas-Maraver.book.2014} and the non-integrable $\varphi^4$ model \cite{Kevrekidis.book.2019}. The integrability of the sine-Gordon model makes it possible to construct exact multisoliton solutions. In the $\varphi^4$ model, this is not possible. Nevertheless, the $\varphi^4$ model has many physical applications and a rich history of studying the dynamical properties of its kink solutions, starting from the 70s of the last century, see, e.g., \cite{Kudryavtsev.JETPLett.1975,Ablowitz.SIAM_JAM.1979,Campbell.PhysD.1983.phi4,Belova.PhysD.1988,Goodman.SIAM_JADS.2005,Goodman.PRL.2007,Takyi.PRD.2016}.

Recently, there has been great interest in the search for kink solutions in various models \cite{Khare.JPA.2019,Khare.PS.2019,Khare.JPA.2020,Kumar.IJMPB.2021,Gani.PRD.2020.explicit,Blinov.JPCS.2020.phi8,Blinov.CSF.2022}, as well as in the study of the dynamics of kink-(anti)kink and multi-kink interactions \cite{Dorey.PRL.2011,Gani.PRD.2014,Adam.PRD.2022.phi6,Gani.JHEP.2015,Gani.PRE.1999,Gani.EPJC.2018.dsg,Belendryasova.JPCS.2019.dsg,Belendryasova.PLB.2021,Christov.CNSNS.2021,Peyrard.PhysD.1983.msG,Bazeia.EPJC.2018.sinh,Bazeia.JPCS.2017.sinh,Moradi.CSF.2022,Demirkaya.JHEP.2017.phi6_cc,Mohammadi.CNSNS.2021,Mohammadi.CSF.2022.B_phi4_model,Alonso-Izquierdo.PRD.2021,Alonso-Izquierdo.CNSNS.2022.kinks_and_wobblers,Dmitriev.PRE.2008,Moradi.JHEP.2017,Moradi.EPJB.2018,Gani.EPJC.2019.dsg,Gani.EPJC.2021.exotic_phi8,Gonzalez.PLA.1989.long-range,Guerrero.PLA.1998.long-range,Gomes.PRD.2012,Radomskiy.JPCS.2017,Belendryasova.CNSNS.2019,Christov.PRL.2019,Christov.PRD.2019,Manton.JPA.2019,Amado.EPJC.2020,dOrnellas.JPC.2020,Campos.PLB.2021}. In particular, the following results were obtained:
\begin{itemize}
    \item families of logarithmic potentials were studied \cite{Khare.PS.2019,Khare.JPA.2020,Kumar.IJMPB.2021};
    \item exact formulas for the kink solutions in the polynomial models were obtained \cite{Gani.PRD.2020.explicit,Blinov.JPCS.2020.phi8,Blinov.CSF.2022};
    \item kink-(anti)kink collisions were studied in various models --- the $\varphi^6$ model \cite{Dorey.PRL.2011,Gani.PRD.2014,Adam.PRD.2022.phi6}, the $\varphi^8$ model \cite{Gani.JHEP.2015}, the double sine-Gordon model \cite{Gani.PRE.1999,Gani.EPJC.2018.dsg,Belendryasova.JPCS.2019.dsg}, the logarithmic model \cite{Belendryasova.PLB.2021}, models with higher-order polynomial potentials \cite{Christov.CNSNS.2021}, and non-polynomial models \cite{Peyrard.PhysD.1983.msG,Bazeia.EPJC.2018.sinh,Bazeia.JPCS.2017.sinh,Moradi.CSF.2022};
    \item some exotic models were considered \cite{Demirkaya.JHEP.2017.phi6_cc,Mohammadi.CNSNS.2021,Mohammadi.CSF.2022.B_phi4_model} as well as scattering of wobbling kinks \cite{Alonso-Izquierdo.PRD.2021,Alonso-Izquierdo.CNSNS.2022.kinks_and_wobblers};
    \item multi-kink collisions in various models were studied \cite{Dmitriev.PRE.2008,Moradi.JHEP.2017,Moradi.EPJB.2018,Gani.EPJC.2019.dsg,Gani.EPJC.2021.exotic_phi8}.
\end{itemize}

Noteworthy that, in addition to kinks with exponential asymptotics (which, in particular, are the kinks of the above-mentioned sine-Gordon and $\varphi^4$ models), kinks with other asymptotic behavior are studied. In particular, in papers by A.~Khare, A.~Saxena and P.~Kumar, kink solutions are constructed with super-exponential \cite{Kumar.IJMPB.2021} and super-super-exponential asymptotics \cite{Khare.PS.2019}, as well as with asymptotic behavior of the power-tower type \cite{Khare.JPA.2020}. New results have been obtained for kinks with power-law asymptotics \cite{Khare.JPA.2019,Gonzalez.PLA.1989.long-range,Guerrero.PLA.1998.long-range,Gomes.PRD.2012,Radomskiy.JPCS.2017,Belendryasova.CNSNS.2019,Christov.PRL.2019,Christov.PRD.2019,Manton.JPA.2019,Amado.EPJC.2020,dOrnellas.JPC.2020,Campos.PLB.2021}. Many papers are devoted to the study of the interaction forces of kinks having power-law tails \cite{Gomes.PRD.2012,Radomskiy.JPCS.2017,Belendryasova.CNSNS.2019,Christov.PRL.2019,Manton.JPA.2019,Amado.EPJC.2020,dOrnellas.JPC.2020,Campos.PLB.2021}, as well as to the dynamics of collisions of such solitons \cite{Belendryasova.CNSNS.2019,Christov.CNSNS.2021}.

Among the models whose kink solutions are being actively studied, models with polynomial potentials occupy a special place. In particular, new kink solutions have been found for the $\varphi^8$ model \cite{Gani.PRD.2020.explicit}, kink-antikink and multi-kink collisions in the $\varphi^8$ model have been studied \cite{Gani.JHEP.2015,Christov.CNSNS.2021,Belendryasova.CNSNS.2019}. The scattering of kinks with power-law asymptotics in the $\varphi^8$, $\varphi^{10}$ and $\varphi^{12}$ models was studied \cite{Christov.CNSNS.2021}.

In this paper, we consider the $\varphi^{12}$ kinks having exponential asymptotics. Based on explicit formulas for kink solutions, the properties of kinks are studied in detail, and various exotic processes in kink collisions are investigated.

Our paper is organized as follows. In Section \ref{sec:Model}, we introduce the $\varphi^{12}$ field-theoretic model, give explicit formula for its kink solutions, present some basic properties of these kinks. In Section \ref{sec:Force}, we obtain asymptotic estimates for the interaction forces of a kink and an antikink located at a large distance from each other. Section \ref{sec:Scattering} is devoted to studying the processes of kink-antikink and antikink-kink collisions in various topological sectors of the model. Finally, in Section \ref{sec:Conclusion}, we briefly summarize and discuss prospects for further research.

\section{The \texorpdfstring{$\varphi^{12}$}{pdfbookmark} model}
\label{sec:Model}

\begin{figure}[t!]
    \centering
    \subfigure[]{\includegraphics[width=0.45 \textwidth, height=0.25 \textheight]{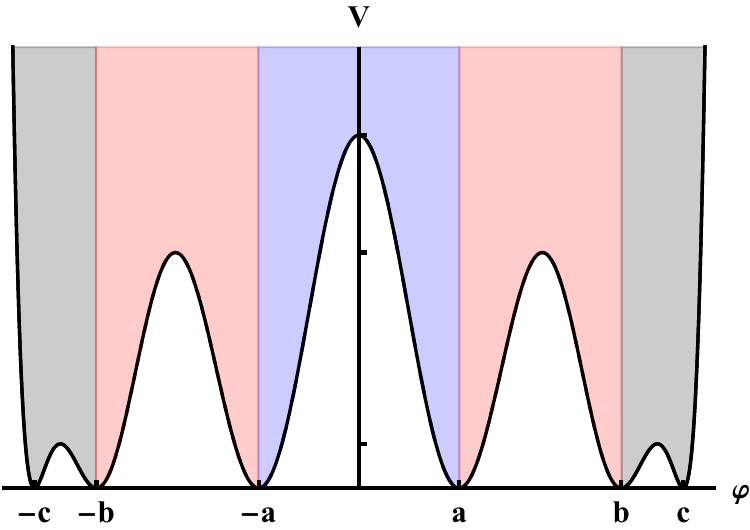}\label{fig:Potential}}
    \subfigure[]{\includegraphics[width=0.45\textwidth, height=0.25 \textheight]{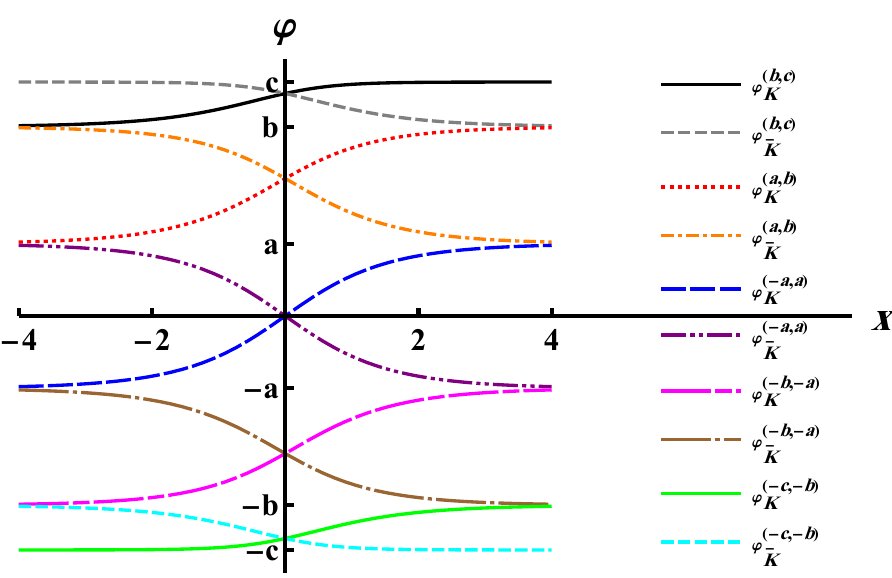}\label{fig:Solutions}}
    \caption{(a) Potential \eqref{eq:potential} for the parameters \eqref{eq:parameters}. Different colors above the curve indicate different topological sectors; mirror-symmetric sectors are shown by the same color. (b) Kink solutions \eqref{eq:all_kinks}.
    }
    \label{fig:SolutionsPotential}
\end{figure}

We consider a field-theoretic model with a real scalar field $\varphi(x,t)$, which evolves in $(1+1)$-dimensional space-time according to the Lagrangian density
\begin{equation}\label{eq:Largangian}
    \mathcal{L} = \frac{1}{2}\left(\frac{\partial \varphi}{\partial t}\right)^2 - \frac{1}{2}\left(\frac{\partial \varphi}{\partial x}\right)^2 - V(\varphi),
\end{equation}
with the following polynomial potential of the twelfth degree
\begin{equation}\label{eq:potential}
    V(\varphi) = \lambda^2 \left(\varphi^2-a^2\right)^2 \left(\varphi^2-b^2\right)^2 \left(\varphi^2-c^2\right)^2,
\end{equation}
where $0<a<b<c$ and $\lambda>0$ are real parameters. This potential has six degenerate minima at $\varphi = \pm a$, $\pm b$, $\pm c$, that split the interval $-c\le\varphi\le c$ into five {\it topological sectors}: $(-c,-b)$, $(-b,-a)$, $(-a,a)$, $(a,b)$, and $(b,c)$, see Fig.~\ref{fig:Potential}.
According to this, ten kink solutions (five kinks ($K$) and five antikinks ($\bar{K}$), see Fig.~\ref{fig:Solutions}) exist in the model. Due to the symmetry of the potential, asymmetric sectors located in the region of negative field values ($(-c,-b)$ and $(-b,-a)$) are mirror-symmetric to those located in the region of positive field values ($(a,b)$ and $(b,c)$).

Lagrangian \eqref{eq:Largangian} leads to the equation of motion
\begin{equation}\label{eq:EOM}
    \frac{\partial^2\varphi}{\partial t^2} - \frac{\partial^2\varphi}{\partial x^2} + \frac{dV}{d\varphi} = 0,
\end{equation}
and the energy functional
\begin{equation}\label{eq:energy}
     E[\varphi] = \int\limits_{-\infty}^{\infty}\left[\frac{1}{2} \left( \frac{\partial\varphi}{\partial t} \right)^2 + \frac{1}{2} \left( \frac{\partial\varphi}{\partial x} \right)^2 + V(\varphi)\right]dx.
\end{equation}
In the case of static kink solutions, Eq.~\eqref{eq:EOM} can be reduced to the first order ordinary differential equation
\begin{equation}\label{eq:BPS}
    \frac{d\varphi}{dx} =\pm \sqrt{2V(\varphi)},
\end{equation}
see, e.g., \cite[Sec.~2]{Gani.JHEP.2015}. For particular choice of the parameters
\begin{equation}\label{eq:parameters}
    a=\frac{\sqrt{5}-1}{4}, \quad b=\frac{\sqrt{5}+1}{4}, \quad c=1, \quad \lambda=\frac{8\sqrt{2}}{5},
\end{equation}
Eq.~\eqref{eq:BPS} can be integrated taking into account appropriate boundary conditions: the field must approach two neighboring vacua at $x\to\pm\infty$. This leads to the explicit formula for all kink solutions of Eq.~\eqref{eq:BPS} (see, e.g., Ref.~\cite{Bazeia.PRD.2006} for more details):
\begin{eqnarray}\label{eq:all_kinks}
    \varphi_{\scriptsize\mbox{K}\mbox{($\bar{\rm K}$)}}^{}(x) = \cos \left(\frac{1}{5} \arccos\left(\tanh x\right)+\frac{\pi s}{5}\right), \quad \mbox{where} \quad s=0, 1, ..., 9.
\end{eqnarray}
In Eq.~\eqref{eq:all_kinks}, $s$ runs through any ten consecutive integer values, that we have taken from 0 to 9. This equation gives all ten kinks and antikinks of the model with the potential \eqref{eq:potential}, see Fig.~\ref{fig:Solutions}.

Mass of a kink can be found either by substituting Eq.~\eqref{eq:all_kinks} into Eq.~\eqref{eq:energy} or by using {\it superpotential} (also sometimes called {\it prepotential}) $W(\varphi)$, which is related to the potential $V(\varphi)$ by
\begin{equation}\label{eq:potential_and_superpotential}
    V(\varphi) = \frac{1}{2}\left(\frac{dW}{d\varphi}\right)^2,
\end{equation}
see, e.g., \cite[Sec.~2]{Gani.JHEP.2015}. For the potential \eqref{eq:potential}, the superpotential can be written as
\begin{equation}\label{eq:superpotential}
    W(\varphi) = \lambda \sqrt{2} \left( \frac{1}{7} \cdot \varphi ^7 
    - \frac{a^2 + b^2 + c^2}{5} \cdot \varphi ^5
    + \frac{a^2 b^2 + a^2 c^2 + b^2 c^2}{3} \cdot \varphi ^3 - a^2 b^2 c^2 \cdot \varphi \right). 
\end{equation}
Masses of all kinks of the model \eqref{eq:potential} then look like
\begin{equation}
    M_{\scriptsize\mbox{K}}^{(-a,a)} = \lambda \sqrt{2} \left( \frac{4}{35} \cdot a^7  - \frac{4}{15} \left(b^2 + c^2  \right) \cdot a^5 + \frac{4}{3} c^2 b^2 \cdot a^3 \right),
\end{equation}
\begin{equation}
   M_{\scriptsize\mbox{K}}^{(a,b)} = \lambda \sqrt{2} \left( a - b \right)^3 \left(3 ( a^4+b^4) + 9ab(a^2+b^2) + 11 a^2b^2 - 7c^2 (a^2 + b^2) - 21 abc^2 \right),
\end{equation}
\begin{equation}
    M_{\scriptsize\mbox{K}}^{(b,c)} = \lambda \sqrt{2} \left( c - b \right)^3 \left(3 (b^4 + c^4) + 9bc(b^2+c^2) + 11 b^2c^2 - 7a^2 (b^2 + c^2) - 21 a^2bc \right).
\end{equation}
Masses of kinks \eqref{eq:all_kinks} are given in Table~\ref{tab:Table1}.
\begin{table}[t!]
    \caption{Mass properties of kinks and antikinks \eqref{eq:all_kinks}.}
    \begin{center}
    \begin{tabular}{c|c|c|c}
    \hline \hline
    topological sector &  $s$ in Eq.~\eqref{eq:all_kinks}  &  symmetry of the (anti)kink & mass of the (anti)kink \\
    \hline
    $(-a,a)$    &   2, 7  &  symmetric  & $\frac{25 \sqrt{5}+109}{2100} \approx 0.0785$ \\
    \hline
    $(a,b)$, $(-b,-a)$   & \phantom{V}  1, 3, 6, 8 \phantom{V}  &  asymmetric & $\frac{109}{2100}\approx0.0519$\\
    \hline
    $(b,c)$, $(-c,-b)$   &   0, 4, 5, 9  &  asymmetric & $\frac{93-25 \sqrt{5}}{4200}\approx 0.0088$\\
    \hline
    \end{tabular}
    \end{center}
    \label{tab:Table1}
\end{table}
The symmetric kink (antikink) in the topological sector $(-a,a)$ is heavier than the asymmetric ones in the sectors $(a,b)$ and $(b,c)$. Besides that, the kink in the sector $(a,b)$ is significantly heavier than in the sector $(b,c)$, $ M_{\scriptsize\mbox{K}}^{(b,c)}< M_{\scriptsize\mbox{K}}^{(a,b)}< M_{\scriptsize\mbox{K}}^{(-a,a)}$. Looking ahead, we can assume that due to such a difference in masses, some non-trivial processes in kink-antikink collisions can occur, see, e.g., \cite{Gani.EPJC.2021.exotic_phi8}.

The masses of small perturbations around the vacua are defined by the second derivative of the potential:
\begin{equation}
    m_a^{} = \left.\frac{d^2V}{d\varphi^2}\right|_{\varphi=a} = 8 a^2 \lambda^2 \left(a^2-b^2\right)^2 \left(a^2-c^2\right)^2,
\end{equation}
\begin{equation}
    m_b^{} = \left.\frac{d^2V}{d\varphi^2}\right|_{\varphi=b} = 8 b^2 \lambda^2 \left(a^2-b^2\right)^2 \left(b^2-c^2\right)^2,
\end{equation}
\begin{equation}
    m_c^{} = \left.\frac{d^2V}{d\varphi^2}\right|_{\varphi=c} = 8 c^2 \lambda^2 \left(a^2-c^2\right)^2 \left(b^2-c^2\right)^2.
\end{equation}
For the constants \eqref{eq:parameters} we get $m_a^{}=m_b^{}=1$ and $m_c^{}=4$.

Below in this paper, almost everywhere in formulas we will use the specific values of the constants \eqref{eq:parameters}.


\section{Force of interaction between kink and antikink}
\label{sec:Force}

Due to the non-integrability of the field-theoretic model under consideration, it does not have multisoliton solutions \cite{Manton.book.2004,Shnir.book.2018}. However, well-separated kink and antikink satisfy the equation of motion with exponential (in distance) accuracy. The nonlinearity of the model leads to the fact that an attractive force arises between kink and antikink.

In this section, in order to estimate the forces between kink and antikink in all topological sectors of the model, we use an asymptotic method described, e.g., in \cite[Sec.~5.2]{Manton.book.2004}.
The main idea of the method is as follows. The momentum of a field configuration $\varphi(x,t)$ on the semi-infinite interval $(-\infty,\beta]$ looks like
\begin{equation}\label{eq:momentum}
    P = - \int_{-\infty}^\beta \frac{\partial\varphi}{\partial t} \frac{\partial\varphi}{\partial x} dx.
\end{equation}
Hence, the force acting on this interval is
\begin{equation}\label{intervalForce}
    F = \frac{\partial P}{\partial t} = - \int_{-\infty}^{\beta} \left(\frac{\partial^2\varphi}{\partial t^2}\frac{\partial\varphi}{\partial x}+\frac{\partial\varphi}{\partial t} \frac{\partial^2\varphi}{\partial x \partial t}\right) dx.
\end{equation}
Simplifying the above formula by using Eq.~\eqref{eq:EOM}, we obtain
\begin{equation}\label{intervalForce3}
    F = \left.\left[-\frac{1}{2}\left(\frac{\partial\varphi}{\partial t}\right)^2 -\frac{1}{2}\left(\frac{\partial\varphi}{\partial x}\right)^2+V(\varphi)\right]\right|^\beta_{-\infty}.
\end{equation}

For example, if we are interested in the force between the static kink and antikink in the topological sector $(-a,a)$, we have to use the field configuration in the form of kink $\varphi_{\scriptsize\mbox{K}}^{(-a,a)}(x)$ and antikink $\varphi_{\scriptsize\mbox{$\bar{\rm K}$}}^{(-a,a)}(x)$ that are centered in the points $x=-X$ and $x=X$, respectively:
\begin{eqnarray}\label{eq:K-aK}
    \varphi_{\scriptsize\mbox{K$\bar{\rm K}$}}^{(-a,a)}(x) &=& \varphi_{\scriptsize\mbox{K}}^{(-a,a)}(x+X) + \varphi_{\scriptsize\mbox{$\bar{\rm K}$}}^{(-a,a)}(x-X) - a.
\end{eqnarray}
We assume that $X\gg 1$ and $-X\ll \beta \ll X$, hence $\varphi_{\scriptsize\mbox{$\bar{\rm K}$}}^{(-a,a)}(x-X) - a$ is exponentially small at $x\le\beta$ and tends to zero as $X$ goes to $+\infty$. Thus, substituting Eq.~\eqref{eq:K-aK} into Eq.~\eqref{intervalForce3} and linearizing up to the first order of $\varphi_{\scriptsize\mbox{$\bar{\rm K}$}}^{(-a,a)}(x-X) - a$, we obtain
\begin{eqnarray}\label{force1Ia}
    F_{\scriptsize\mbox{K}\mbox{$\bar{\rm K}$}}^{(-a,a)} &=& \bigg[
    - \frac{1}{2}\left(\frac{\partial  \varphi_{\scriptsize\mbox{K$\bar{\rm K}$}}^{(-a,a)}(x)}{\partial x}\right)^2 
    + V\left(\varphi_{\scriptsize\mbox{K$\bar{\rm K}$}}^{(-a,a)}(x)\right)\bigg]\bigg|^\beta_{-\infty}
\nonumber \\
    &\cong & \bigg[-\frac{1}{2}\left( \frac{\partial\varphi_{\scriptsize\mbox{K}}^{(-a,a)}(x+X)}{\partial x} \right)^2 - 
    \frac{\partial\varphi_{\scriptsize\mbox{K}}^{(-a,a)}(x+X)}{\partial x}\cdot\frac{\partial\varphi_{\scriptsize\mbox{$\bar{\rm K}$}}^{(-a,a)}(x-X)}{\partial x} + V(\varphi_{\scriptsize\mbox{K}}^{(-a,a)}(x+X))\nonumber\\
    &+&\frac{dV(\varphi_{\scriptsize\mbox{K}}^{(-a,a)}(x+X))}{d\varphi}\cdot\left(\varphi_{\scriptsize\mbox{$\bar{\rm K}$}}^{(-a,a)}(x-X)-a\right)\bigg]\bigg|_{-\infty}^\beta.
\end{eqnarray}
Formula \eqref{force1Ia} gives an estimate of the force acting on a kink centered at the point $x=-X$ from an antikink centered at $x=X$.

Since the kink and antikink satisfy Eqs.~\eqref{eq:EOM} and \eqref{eq:BPS},
we finally have for the force:
\begin{eqnarray}\label{force2Iaa}
    F_{\scriptsize\mbox{K}\mbox{$\bar{\rm K}$}}^{(-a,a)}&\cong &  \bigg[ -\frac{\partial\varphi_{\scriptsize\mbox{K}}^{(-a,a)}(x+X)}{\partial x}\frac{\partial\varphi_{\scriptsize\mbox{$\bar{\rm K}$}}^{(-a,a)}(x-X)}{\partial x} +\left(\varphi_{\scriptsize\mbox{$\bar{\rm K}$}}^{(-a,a)}(x-X)-a\right) \frac{\partial^2\varphi_{\scriptsize\mbox{K}}^{(-a,a)}(x+X)}{\partial x^2 }         \bigg]\bigg|_{-\infty}^\beta.
\end{eqnarray}
The point $\beta$ is far from kink and antikink, and we are dealing with a field configuration, spatial derivatives of which fall off exponentially at spatial infinity, hence we can apply asymptotic forms
\begin{equation}\label{kinkI}
    \varphi_{\scriptsize\mbox{K}}^{(-a,a)}(x+X) = \cos \left(\frac{1}{5} \arccos\left(\tanh (x+X)\right)+\frac{2}{5}\pi\right) \approx a -\frac{1}{5} \sqrt{\frac{\sqrt{5}+5}{2}} e^{-(x+X)}
\end{equation}
and
\begin{equation}\label{antikinkI}
    \varphi_{\scriptsize\mbox{$\bar{\rm K}$}}^{(-a,a)}(x-X) = \cos \left(\frac{1}{5} \arccos\left(\tanh (x-X)\right)+\frac{7}{5} \pi \right) \approx a -\frac{1}{5} \sqrt{\frac{\sqrt{5}+5}{2}} e^{x-X},
\end{equation}
so that Eq.~\eqref{force2Iaa} yields
\begin{equation}\label{force2Ia}
    F_{\scriptsize\mbox{K}\mbox{$\bar{\rm K}$}}^{(-a,a)} \cong
    \frac{\sqrt{5}+5}{25}e^{-R},
\end{equation}
where $R=2X$ is the kink-antikink separation.
As can be seen, there is an attractive force between the kink and antikink, and this force falls off exponentially with distance. Notice that the force is independent of $\beta$, which was used just as an auxiliary parameter.

\begin{figure}[t] 
 \centering 
\subfigure[]{\includegraphics[width=0.45
 \textwidth]{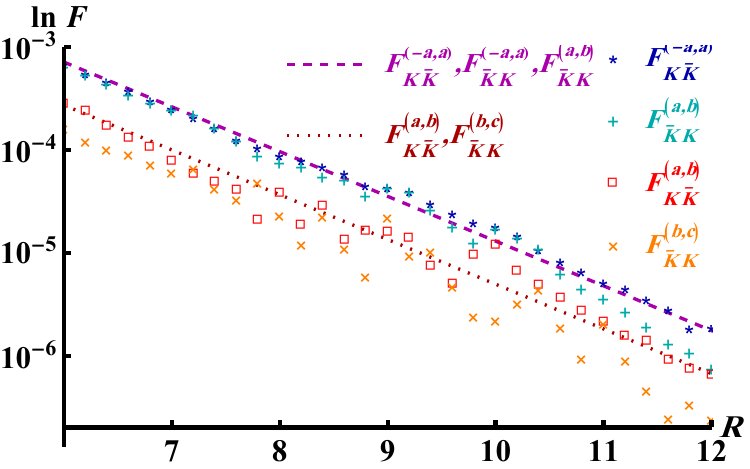}}
 \subfigure[]{\includegraphics[width=0.45
 \textwidth]{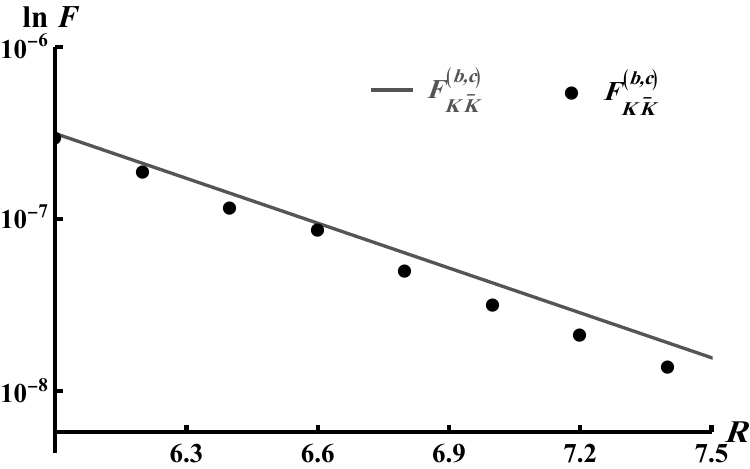}}
 \caption{The kink-antikink and antikink-kink forces as functions of $R$ in various topological sectors. The symbols $*$, $+$, $\square$, $\times$, and $\bullet$ are used for experimental points.
 } 
 \label{fig:Forces} 
 \end{figure}

We also performed the same calculations for kink-antikink (antikink-kink) interaction in the other topological sectors, the results are presented in Table~\ref{tab:ForceEnergyInterval}.
As can be noticed from Table~\ref{tab:ForceEnergyInterval}, the kink-antikink and antikink-kink forces are the same for the symmetric kinks in the topological sector $(-a,a)$. On the contrary, these two forces are different for the asymmetric kinks in the sectors $(a,b)$ and $(b,c)$.
Fig.~\ref{fig:Forces} shows the kink-antikink and antikink-kink forces as functions of $R$ in different topological sectors.
It can also be seen from Table~\ref{tab:ForceEnergyInterval} and Fig.~\ref{fig:Forces} that the above estimate for the antikink-kink force in the sector $(a,b)$ coincides with the kink-antikink force in the sector $(-a,a)$, and the antikink-kink force in the sector $(b,c)$ coincides with the kink-antikink force in the sector $(a,b)$. This is an obvious consequence of the approximation used, within which the force is completely determined by the asymptotic behavior of the kink solution, which, in turn, is completely determined by vacuum the field approaches.

In Fig.~\ref{fig:Forces} we also compare theoretical curves (straight lines on a logarithmic scale) with experimental data. To obtain experimental points, we place the static kink and antikink at some distance. Under the influence of mutual attraction, the solitons begin to move towards each other, and the average force can be estimated from simple mechanical considerations. However, the accuracy of such estimates is not very high, especially for large $R$.

\begin{table}[t!]
    \centering
        \caption{Force between kinks in different topological sectors.}
         \begin{tabular}{r|c|c}
         \hline \hline
  sector\phantom{s} & field configuration & force \\
  \hline 
 \multirow{2}{3em}{$(-a,a)$} & kink-antikink & $\frac{\sqrt{5}+5}{25}e^{-R}$
 \\
                      & antikink-kink & 
 $\frac{\sqrt{5}+5}{25}e^{-R}$
 \\
 \hline
  \multirow{2}{3em}{$(a,b)$} & kink-antikink &
  $\frac{5-\sqrt{5}}{25}e^{-R}$
  \\
                      & antikink-kink &
  $\frac{5+\sqrt{5}}{25}e^{-R}$
  \\
 \hline
  \multirow{2}{3em}{$(b,c)$} & kink-antikink & 
  $\frac{32}{625}e^{-2R}$ \\ 
 & antikink-kink & $\frac{5-\sqrt{5}}{25}e^{-R}$
 \\
 \hline
\end{tabular}
    \label{tab:ForceEnergyInterval}
\end{table}

\section{Kink-antikink and antikink-kink collisions}
\label{sec:Scattering}

Despite the absence of multisoliton solutions in the model under consideration, the physical problem of kink-antikink (antikink-kink) collision can be formulated and solved numerically. The kinks \eqref{eq:all_kinks} have exponential asymptotic behavior at large distances from the region of their localization. This allows us to use as an initial condition a configuration in the form of kink and antikink, separated by a large distance and moving towards each other. This configuration is a solution of equation \eqref{eq:EOM} up to a correction small as an exponential of the kink-antikink distance.

We have performed the numerical simulation of the kink-antikink scattering in different topological sectors. For this purpose, we solved the equation of motion numerically using the fourth-order discretization in space and the Stormer method of integration in time. In our calculations, we used the discretized Eq.~\eqref{eq:EOM}:
\begin{eqnarray}
    &&\frac{d^2 \varphi_{n}}{d t^2} - \frac{1}{h^2} (\varphi_{n-1}-2\varphi_{n}+\varphi_{n+1}) +  \frac{1}{12h^2}(\varphi_{n-2}-4\varphi_{n-1}+6\varphi_{n}-4\varphi_{n+1}+\varphi_{n+2})\nonumber\\
    &&+ \frac{2}{25} \varphi_{n}  \left(768 \varphi_{n} ^{10}-2240 \varphi_{n} ^8+2400 \varphi_{n} ^6-1140 \varphi_{n} ^4+225 \varphi_{n} ^2-13\right) = 0,
\end{eqnarray}
where $n=0,\pm1,\pm2,...$, and the number of nodes is $N=12000$. The temporal and spatial steps used were $\tau=0.005$ and $h=0.025$, respectively. The initial configuration in each case was taken in the form of kink and antikink that are initially located at $x=\pm X_i^{}$ and moving towards each other with velocities $v_i^{}$. For example, in the case of kink-antikink collision in the sector $(-a,a)$, the initial conditions for the numerical solution of the equation of motion were extracted from the configuration
\begin{eqnarray}
    \varphi_{\scriptsize\mbox{K$\bar{\rm K}$}}^{(-a,a)}(x,t) &=& \varphi_{\scriptsize\mbox{K}}^{(-a,a)}\left(\frac{x+X_i^{}-v_i^{}t}{\sqrt{1-v_i^2}}\right) + \varphi_{\scriptsize\mbox{$\bar{\rm K}$}}^{(-a,a)}\left(\frac{x-X_i^{}+v_i^{}t}{\sqrt{1-v_i^2}}\right) - a.
\end{eqnarray}
We now turn to a discussion of the results of numerical experiments.

\subsection{Collision of kink and antikink in the sector \texorpdfstring{$(-a,a)$}{(-a,a)} }\label{sec:case I}

\begin{figure}[t!] 
 \centering 
 \subfigure[\:$v_i^{}=0.10$, $v_f^{} = 0.73$]{\includegraphics[width=0.45
 \textwidth]{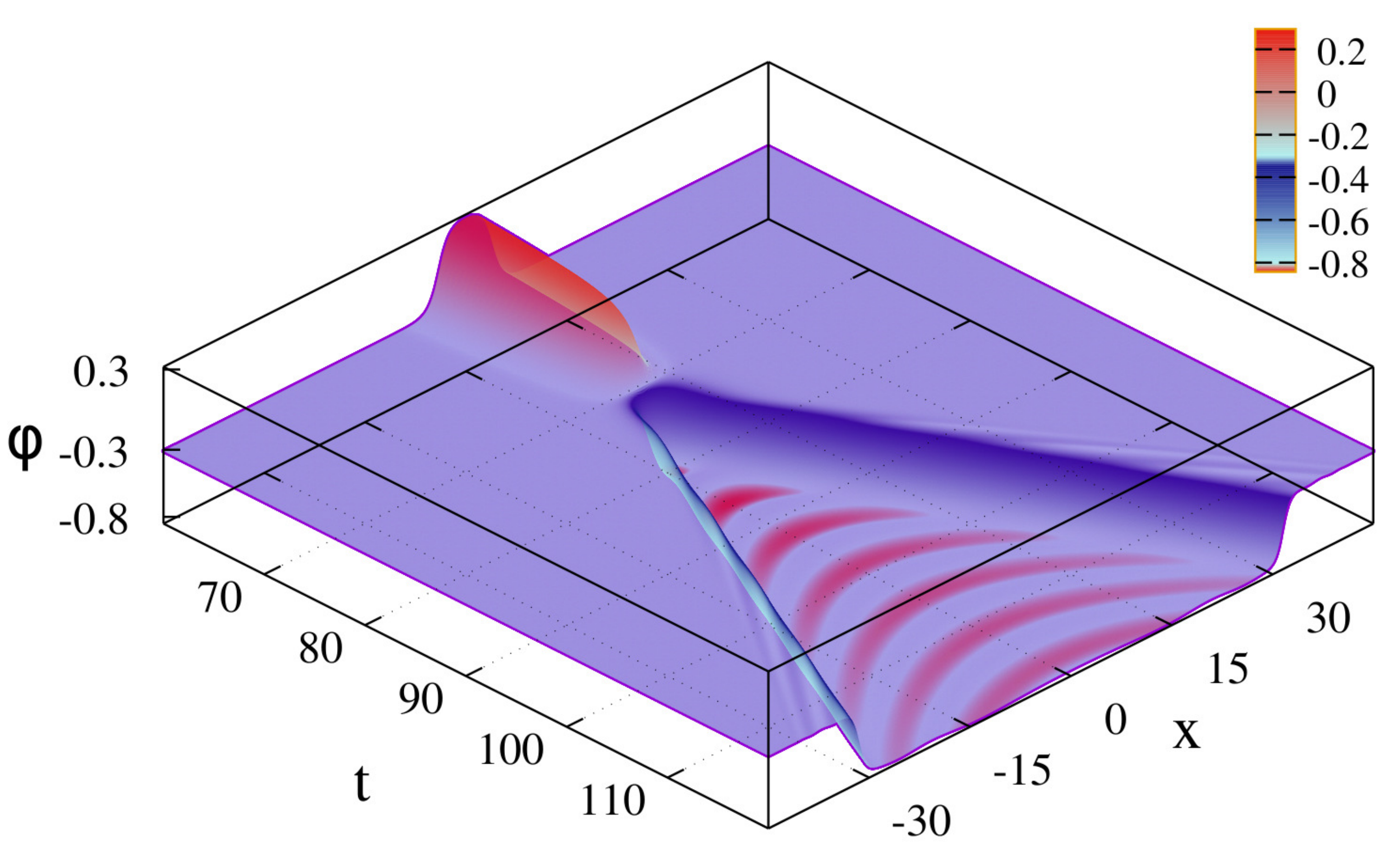}}
\subfigure[\:$v_i^{}=0.90$, $v_f^{} = 0.95$]{\includegraphics[width=0.45
 \textwidth]{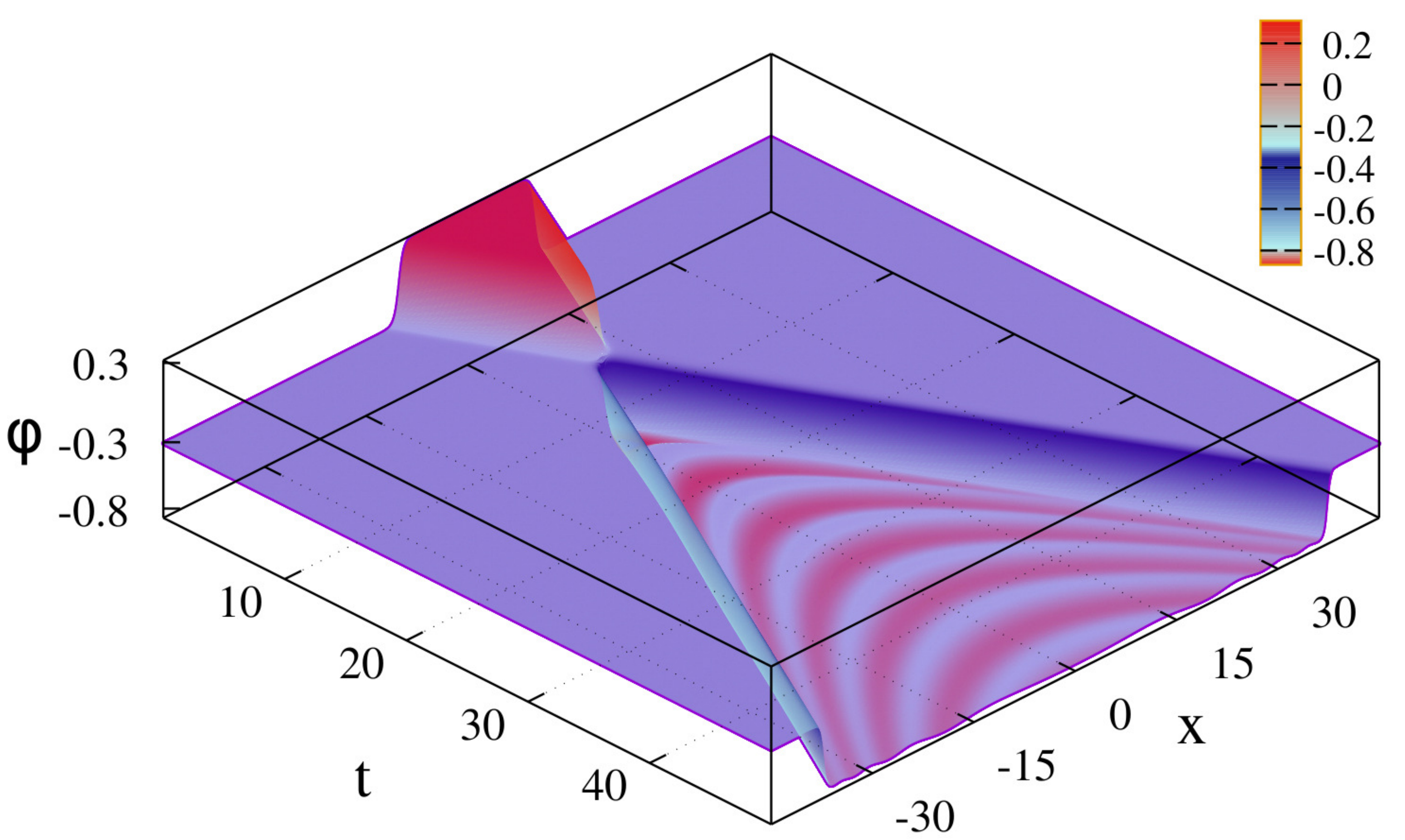}}
\caption{Kink-antikink scattering in the sector $(-a,a)$.
} 
 \label{FigIvariouse velosity} 
\end{figure}

\begin{figure}[t!]
 \centering 
\includegraphics[width=0.45
 \textwidth]{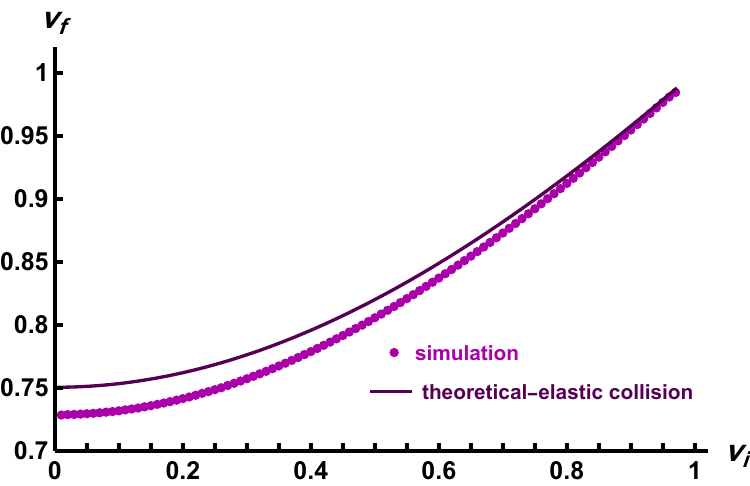}

\caption{The final velocity $v_f^{}$ of the produced kinks in the sector $(-b,-a)$ as a function of the initial velocity $v_i^{}$ of the colliding kinks in the sector $(-a,a)$.}
\label{fig:vfasviTotalRelavesticEnergy}
\end{figure}

The kink and antikink belonging to this topological sector are symmetric with no internal mode (a detailed discussion of finding the vibrational modes of kink and ``kink + antikink'' (or ``antikink + kink'') configurations can be found, e.g., in \cite[Sec.~IV]{Gani.PRD.2020.explicit} and \cite[Sec.~4]{Belendryasova.CNSNS.2019}), and their mass is about $1.5$ times greater than the mass of the kinks in the neighboring sector $(a,b)$ (or $(-b,-a)$), see Table~\ref{tab:Table1}: 
\begin{eqnarray}
    \frac{M_{\scriptsize\mbox{K}}^{(-a,a)}}{M_{\scriptsize\mbox{K}}^{(a,b)}}=\frac{25 \sqrt{5}+109}{109} \approx1.5129.
\end{eqnarray}
The kink and antikink attract each other, and upon collision they can annihilate and form new pair of kink and antikink with smaller masses in the sector $(a,b)$ (or $(-b,-a)$).
This means that there is no critical velocity for kink-antikink (antikink-kink) collision in this sector, and the numerical simulations confirm this scenario, see Fig.~\ref{FigIvariouse velosity}.
For the initial velocities $v_i^{} = 0.10$ and 0.90, the experimentally obtained final velocities are $v_f^{} = 0.73$ and 0.95, respectively. At the same time, if we neglect radiation losses, then for final velocities we obtain the estimates $0.753$ and $0.958$, respectively. A more detailed comparison of the experimental data with the theoretical estimate mentioned above, assuming no energy loss, is shown in Fig.~\ref{fig:vfasviTotalRelavesticEnergy}.

\subsection{Collision of kink and antikink in the sector \texorpdfstring{$(a,b)$}{(a,b)}}
\label{sec:case II}

The kinks in this sector are asymmetric, hence kink-antikink and antikink-kink collisions look different. 

\subsubsection{Kink-antikink collision}\label{sec:2kcaseII}

\begin{figure}[t!] 
 \centering 
\subfigure[\:$v_i^{}=0.4$, $v_f^{}=0$]{\includegraphics[width=0.45
 \textwidth]{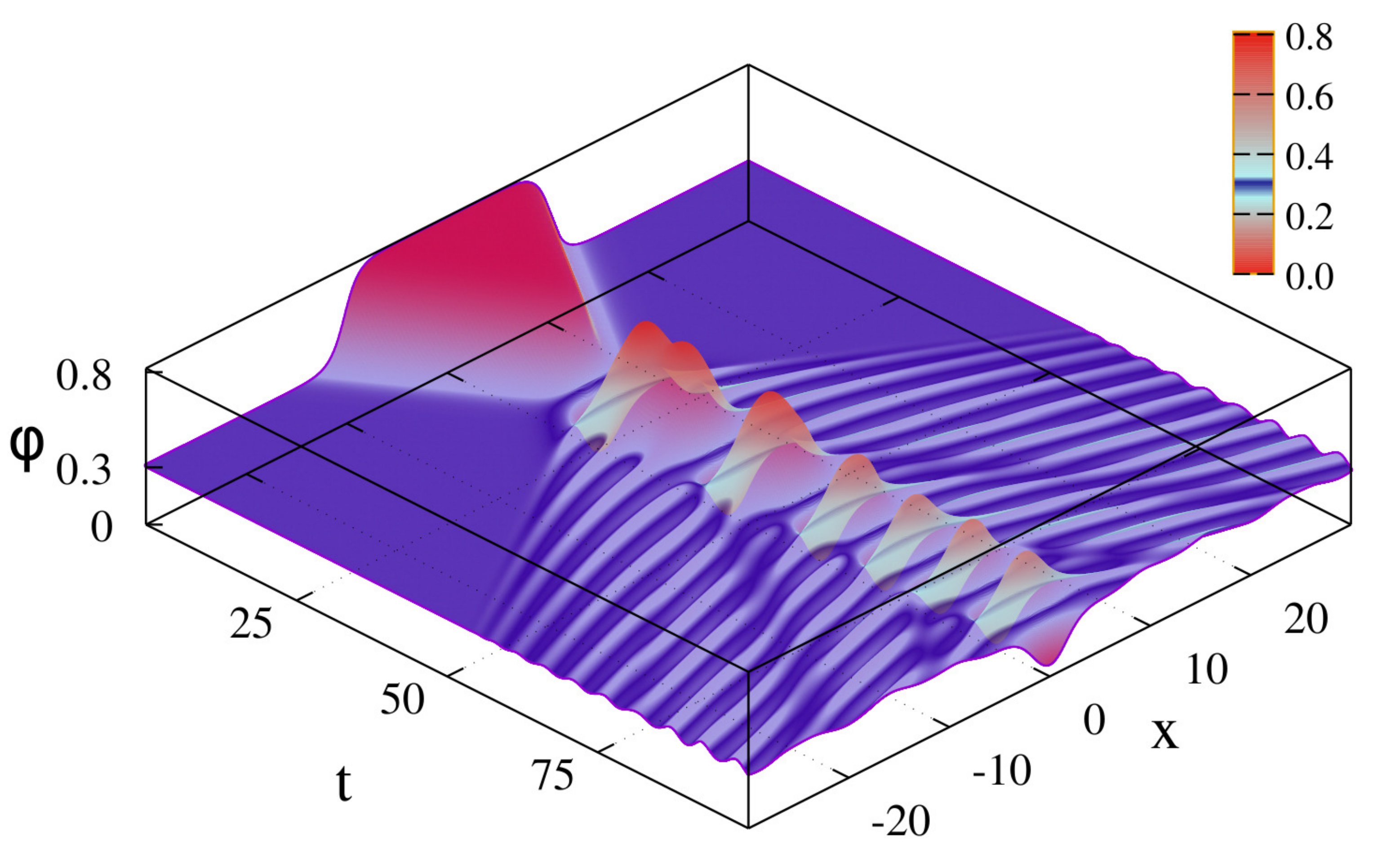}\label{fig:ltvcr}}
\subfigure[\:$v_i^{}=0.5$, $v_f^{}=0.225$]{\includegraphics[width=0.45
 \textwidth]{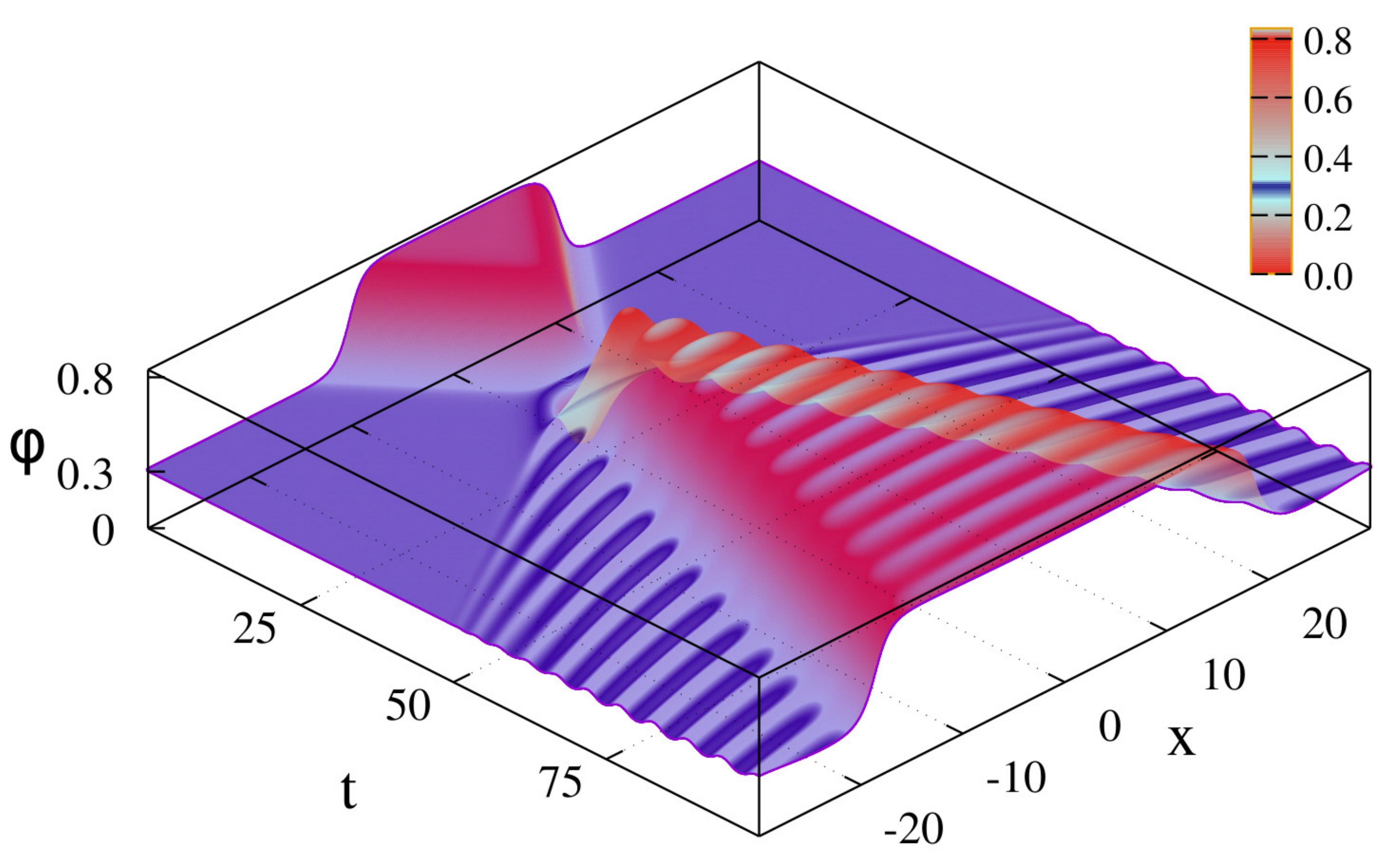}\label{fig:gtvcr}}
\subfigure[\:$v_i^{}=0.7$, $v_f^{}=0.590$]{\includegraphics[width=0.45
 \textwidth]{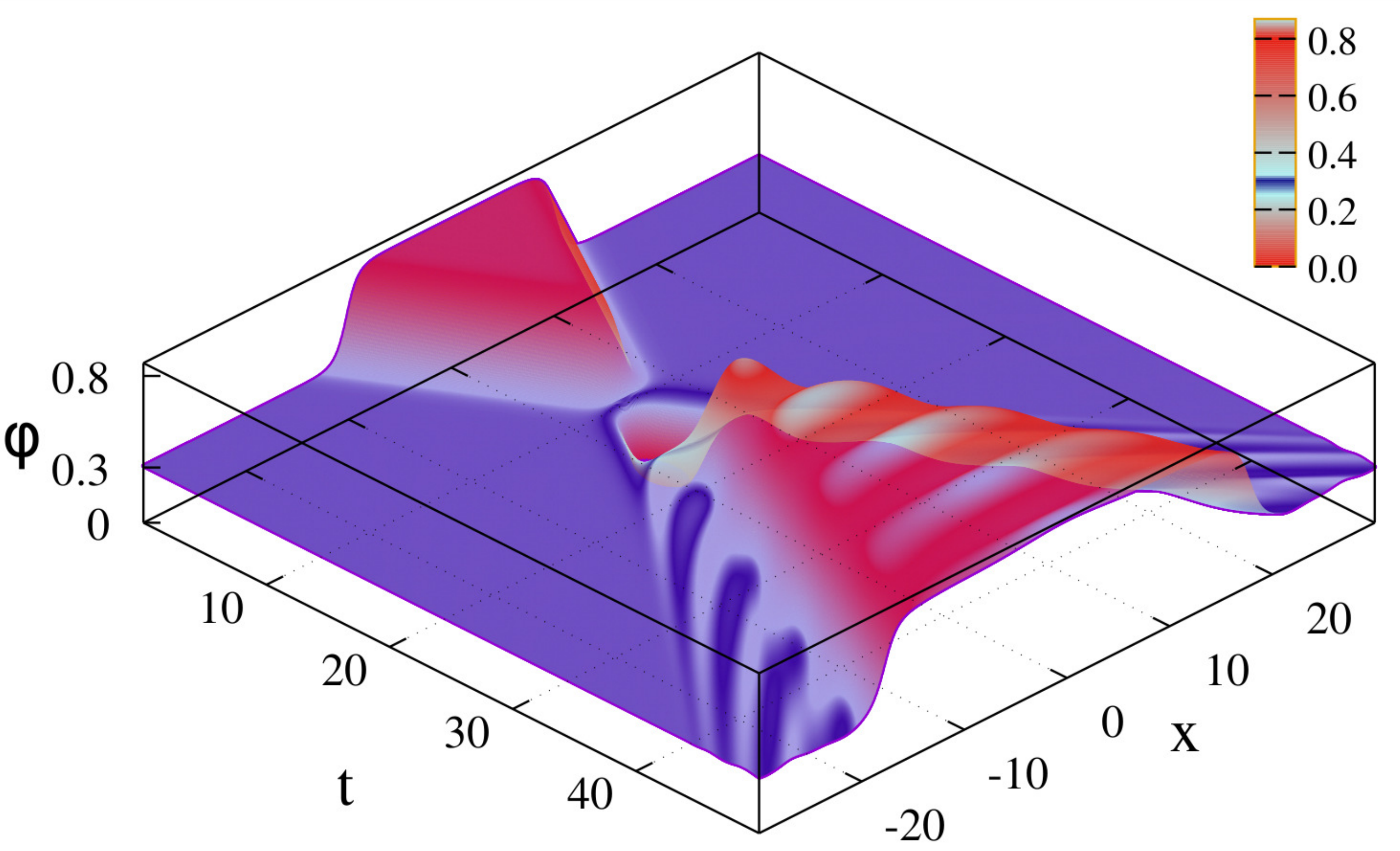}\label{fig:ltvcs}}
\subfigure[\:$v_i^{}=0.8$, $v_f^{}=0.277$
]{\includegraphics[width=0.45
 \textwidth]{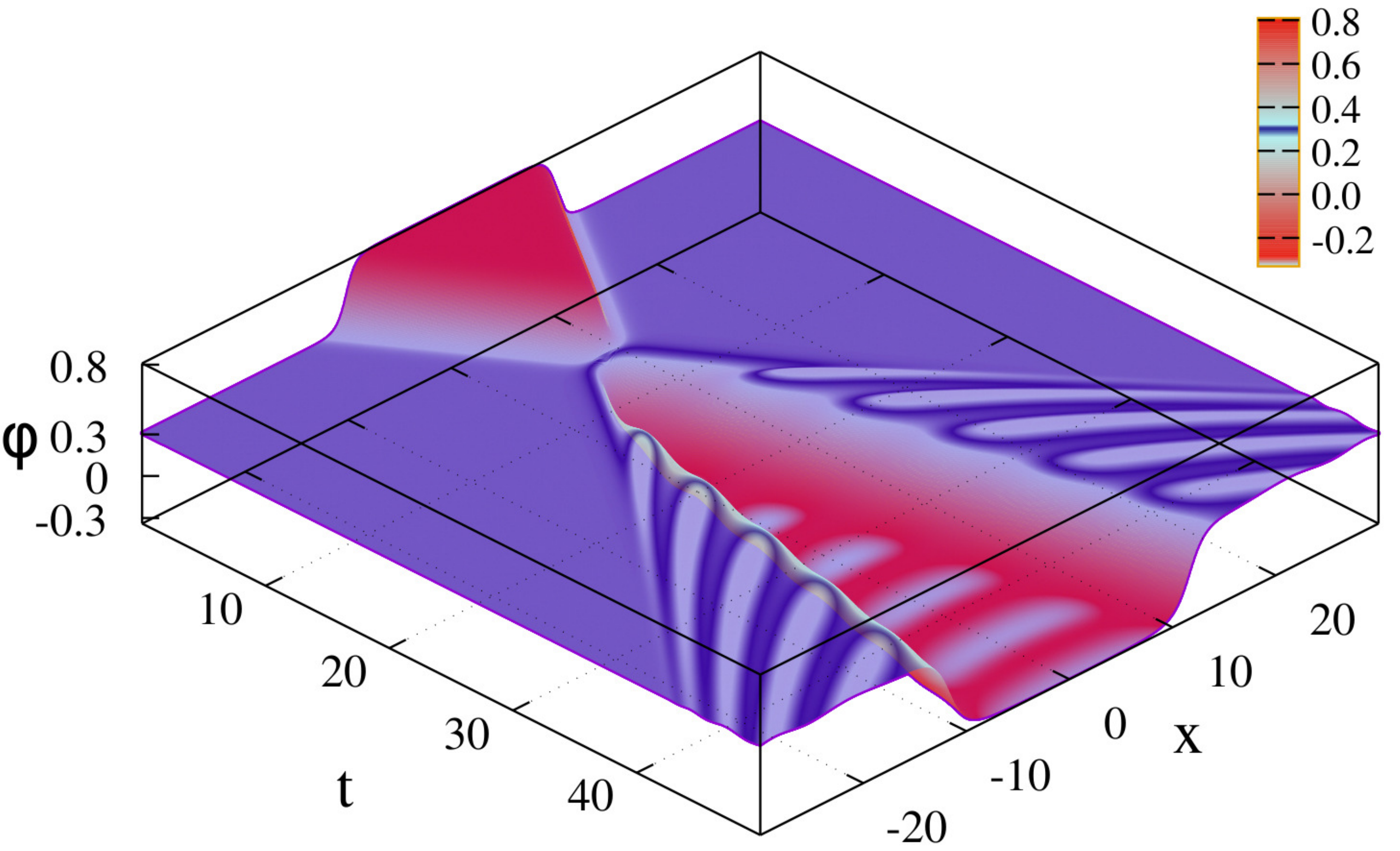}\label{fig:gtvcs}}
\caption{Kink-antikink collision in the sector $(a,b)$. Panel (a) illustrates a bion formation at $v_i^{} < v_{cr}^{}$; plots (b) and (c) show the scattering at $v_{cr}^{} \leq v_i^{} < v_{cs}^{}$; in panel (d), the collision for $v_i^{} \geq v_{cs}^{}$ is presented.}
 \label{fig:kinkantikinkFieldCaseII} 
 \end{figure}

In this case, we observe three different collision scenarios depending on the initial velocity $v_i^{}$: (i) $v_i^{} < v_{cr}^{}$, (ii) $v_{cr}^{} \leq v_i^{} < v_{cs}^{}$, and (iii) $v_i^{} \geq v_{cs}^{}$, where $v_{cr}^{}=0.442$ and ${v_{cs}^{}=0.782}$ (here ``cs'' means ``change sector'').
At $v_i^{} < v_{cr}^{}$, kink and antikink are captured and form a bound state --- a bion, see Fig.~\ref{fig:kinkantikinkFieldCaseII}(a).
If the initial velocity $v_{cr}^{} \leq v_i^{} < v_{cs}^{}$, kink and antikink escape after the collision, and their final velocities are less than the initial, see Figs.~\ref{fig:kinkantikinkFieldCaseII}(b) and \ref{fig:kinkantikinkFieldCaseII}(c). Finally, at the initial velocities more than $v_{cs}^{}$, kink and antikink annihilate and produce an antikink-kink pair in the topological sector $(-a,a)$, see Fig.~\ref{fig:kinkantikinkFieldCaseII}(d).

\begin{figure}[t!]
  \centering \includegraphics[width=0.75\textwidth]{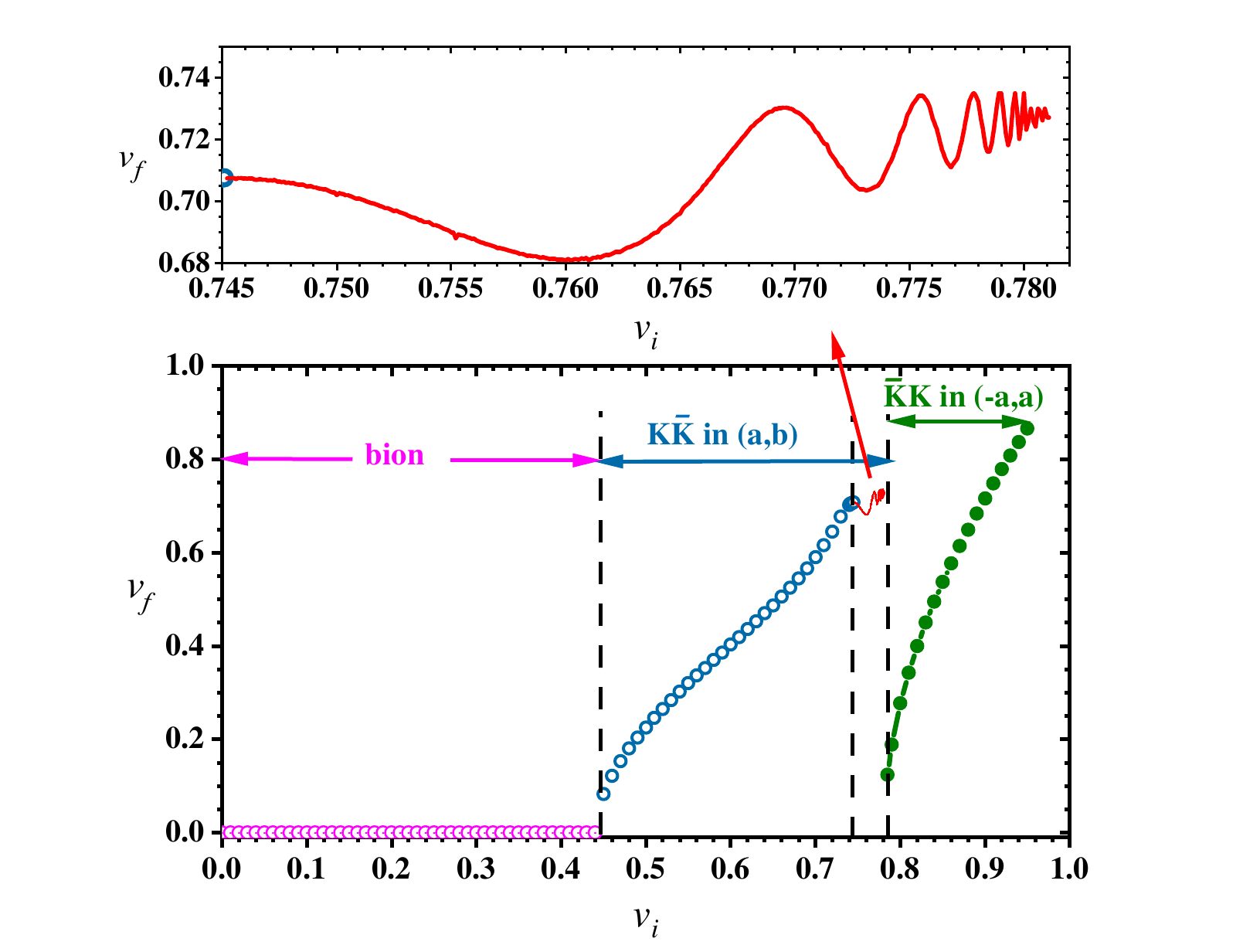}
    \caption{The final velocity as a function of the initial velocity in the kink-antikink collision in the sector $(a,b)$. The pink region corresponds to $v_i^{} < v_{cr}^{}=0.442$, the blue region --- to $v_{cr}^{} \leq v_i^{} < v_{cs}^{}=0.782$, and the green region --- to $v_i^{} \geq v_{cs}^{}$. In a small (shown in red and zoomed on the top panel) interval below $v_{cs}^{}$, the dependence is non-monotonic.}
    \label{fig:VfVi030803}
\end{figure}

\begin{figure}[t!] 
 \centering 
\subfigure[\:$v_i=0.750$, $v_f=0.702$]{\includegraphics[width=0.45
 \textwidth]{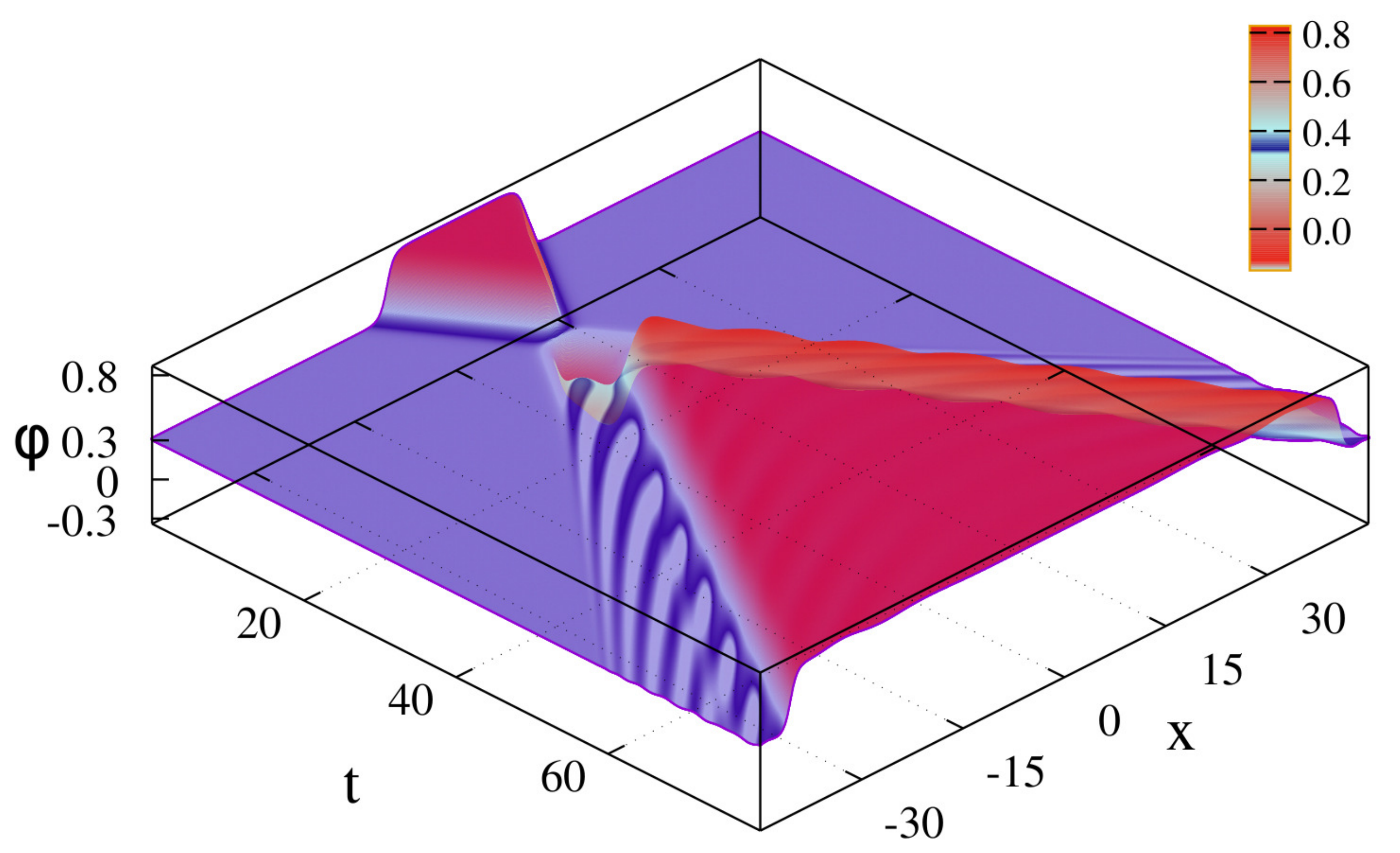}\label{fig:2kP03P08P03Vi0750vf0702}}
\subfigure[\:$v_i=0.765$, $v_f=0.696$]{\includegraphics[width=0.45
 \textwidth]{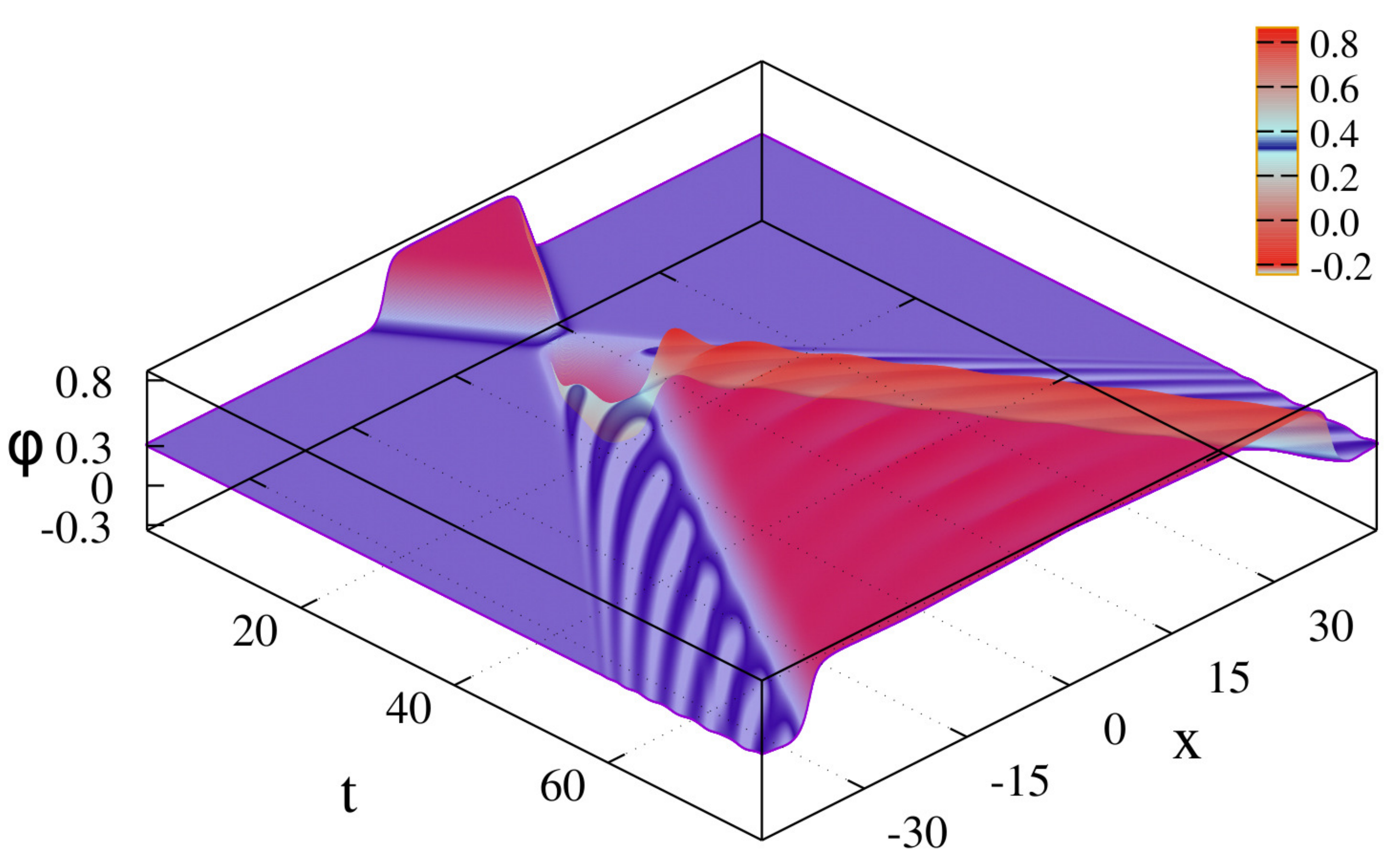}\label{fig:2kP03P08P03Vi0765vf0696}}
\\
\subfigure[\:$v_i=0.775$, $v_f=0.729$]{\includegraphics[width=0.45
 \textwidth]{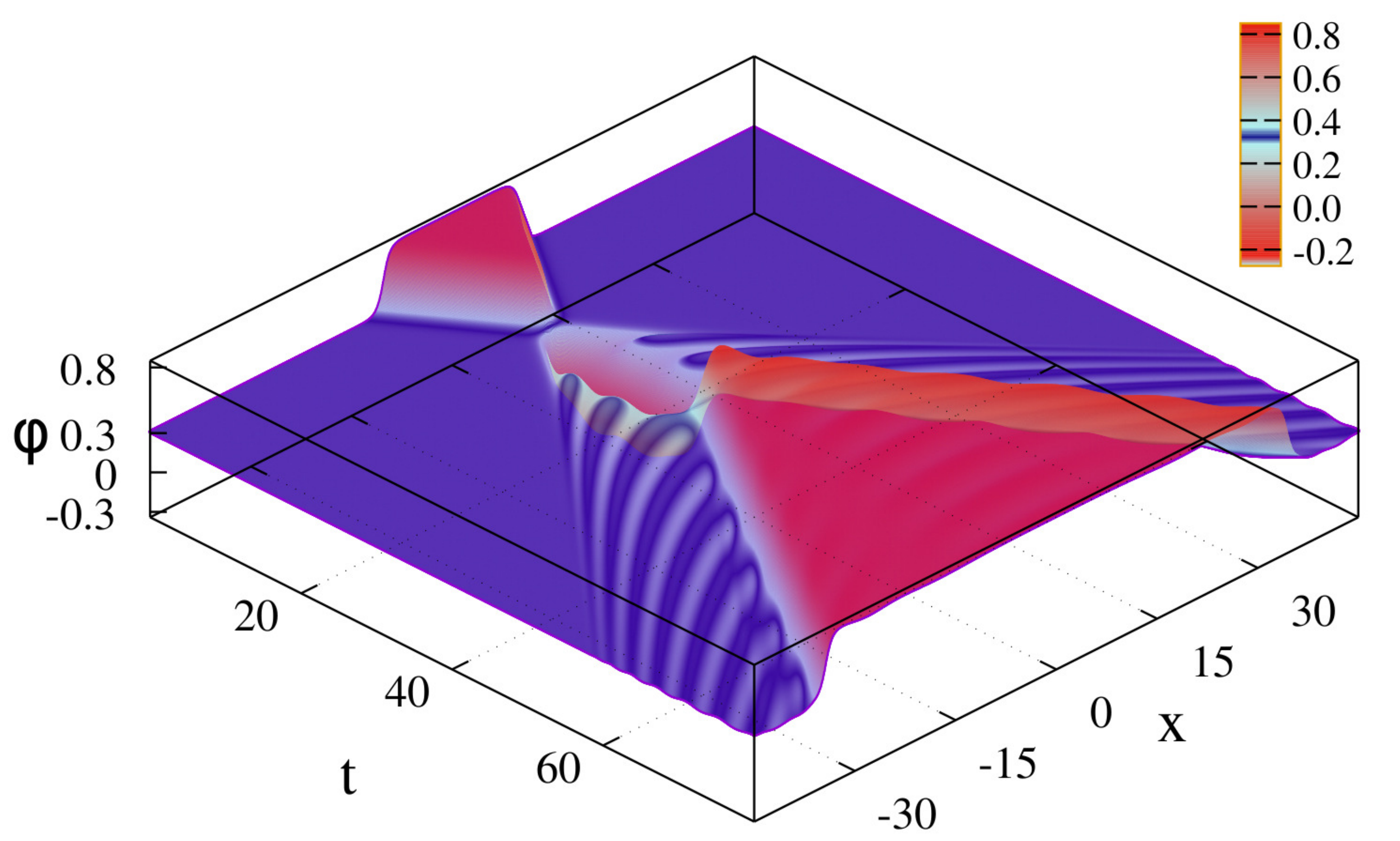}\label{fig:2kP03P08P03Vi0775vf0729}}
 \subfigure[\:$v_i=0.780$, $v_f=0.735$]{\includegraphics[width=0.45
 \textwidth]{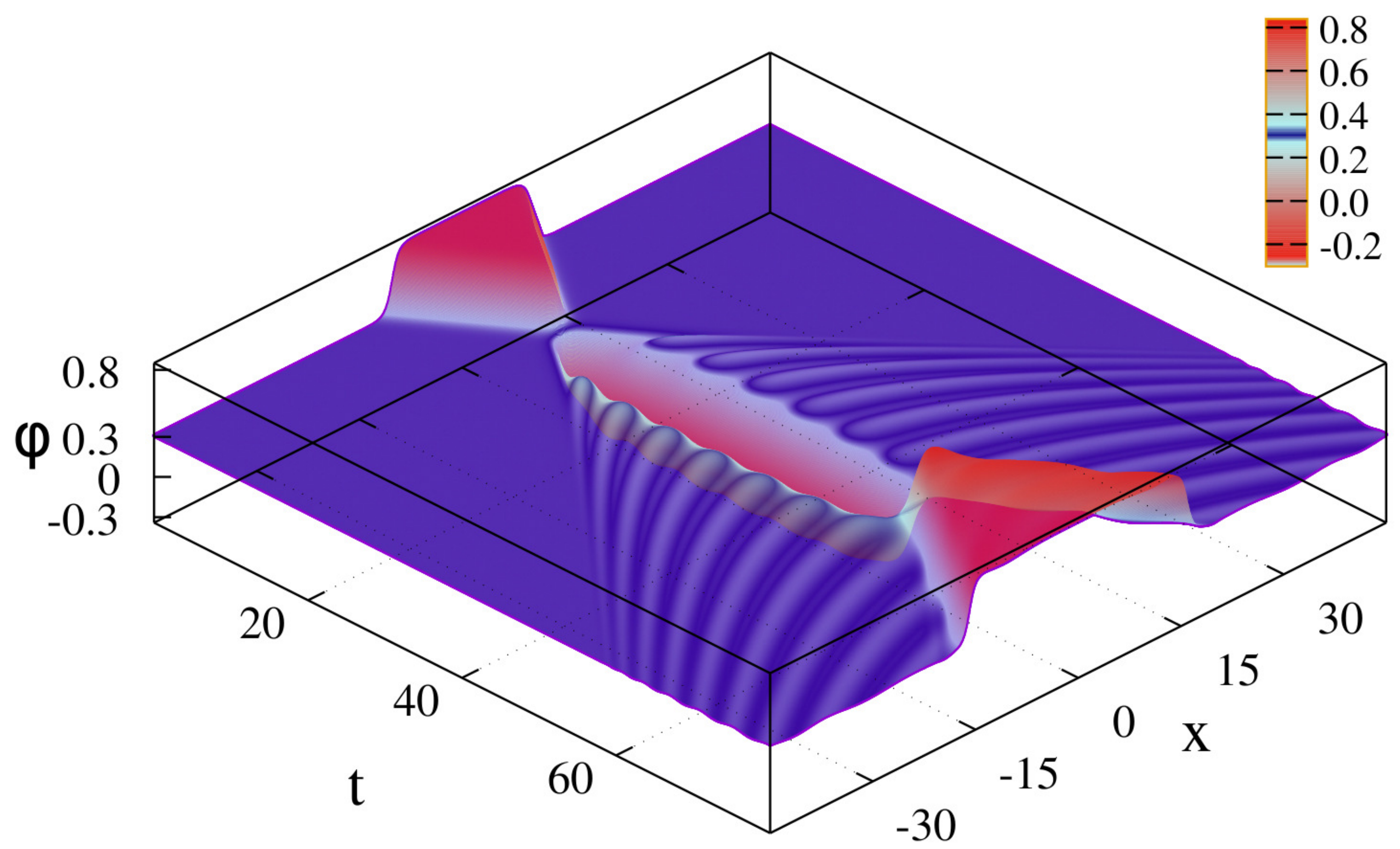}\label{fig:2kP03P08P03Vi0780vf0735}}
  \caption{Kink-antikink collisions in the sector $(a,b)$.} 
 \label{fig:ResonansRegion} 
 \end{figure}

In Fig.~\ref{fig:VfVi030803}
we give the final velocity of kinks as a function of the initial velocity in the sector $(a,b)$. For $v_i^{} < v_{cr}^{}$, we observe the formation of a bion at the collision point, hence the final velocity is assumed to be zero. In the interval $v_{cr}^{} \leq v_i^{} < v_{cs}^{}$, the kinks do not change their topological sector, and the final velocity increases with increasing initial velocity, although at the right end of this interval a kind of resonance behavior is observed. In the range of initial velocities between approximately 0.745 and 0.780, the dependence of $v_f^{}$ on $v_i^{}$ becomes non-monotonic, see top panel of Fig.~\ref{fig:VfVi030803}. The lifetime of the antikink-kink pair in sector $(-a,a)$ becomes longer as the initial velocity increases. Nevertheless, symmetric kinks cannot separate at infinity, and in the final state, the kink and antikink in sector $(a,b)$ are still observed, see Fig.~\ref{fig:ResonansRegion}.
 
Finally, at $v_i^{} \geq v_{cs}^{}$, an antikink-kink pair in the sector $(-a,a)$ is formed in the final state, therefore the final velocity of kinks is much less than the initial velocity.

\subsubsection{Antikink-kink collision}

\begin{figure}[t!] 
 \centering 
 \subfigure[\:$v_i^{}=0.1$, $v_f^{}=0.885$]{\includegraphics[width=0.45
 \textwidth]{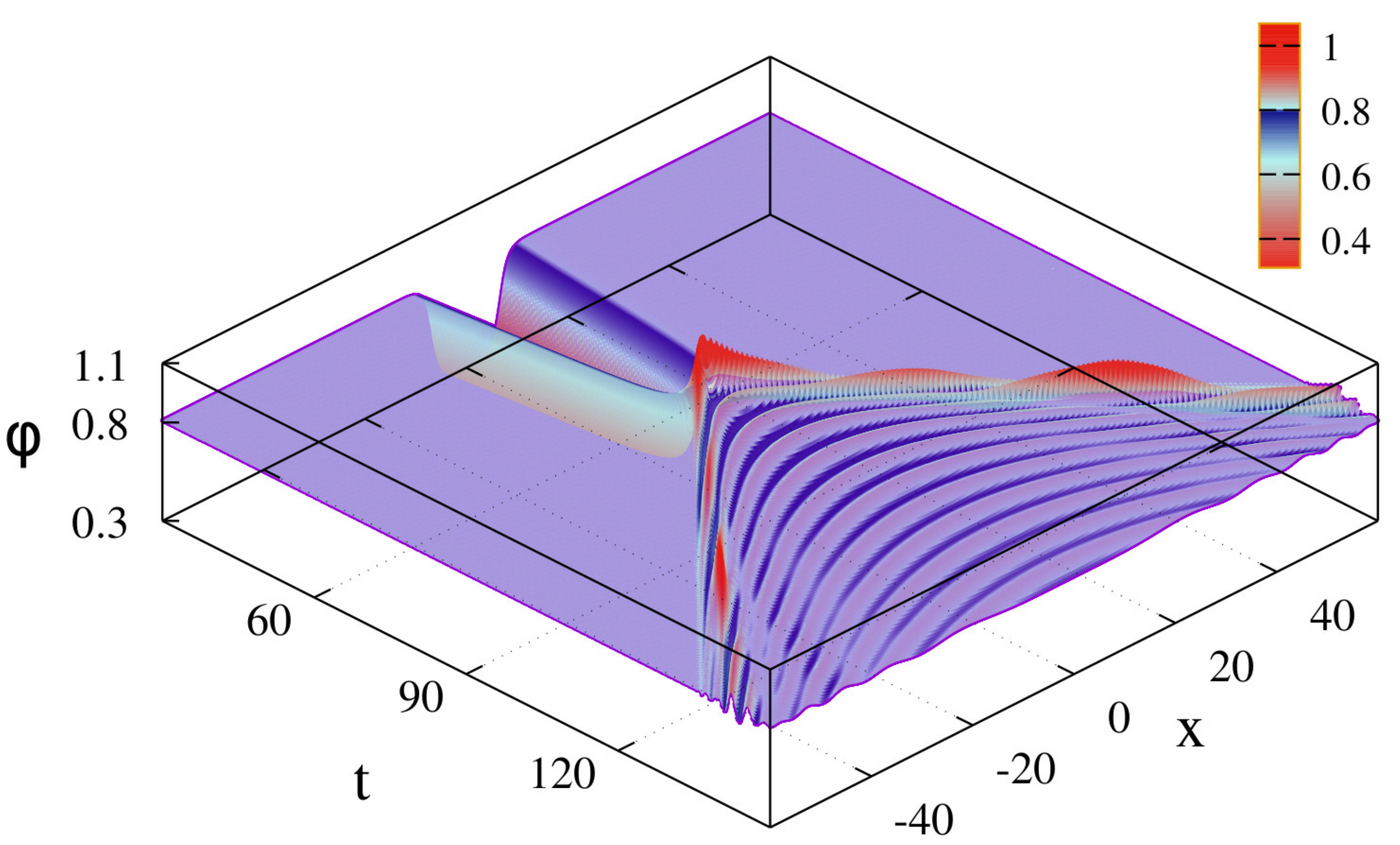}\label{fig:2kP08P03P08vi01vf088504}}
 \subfigure[\:$v_i^{}=0.5$, $v_f^{}=0.924$]{\includegraphics[width=0.45
 \textwidth]{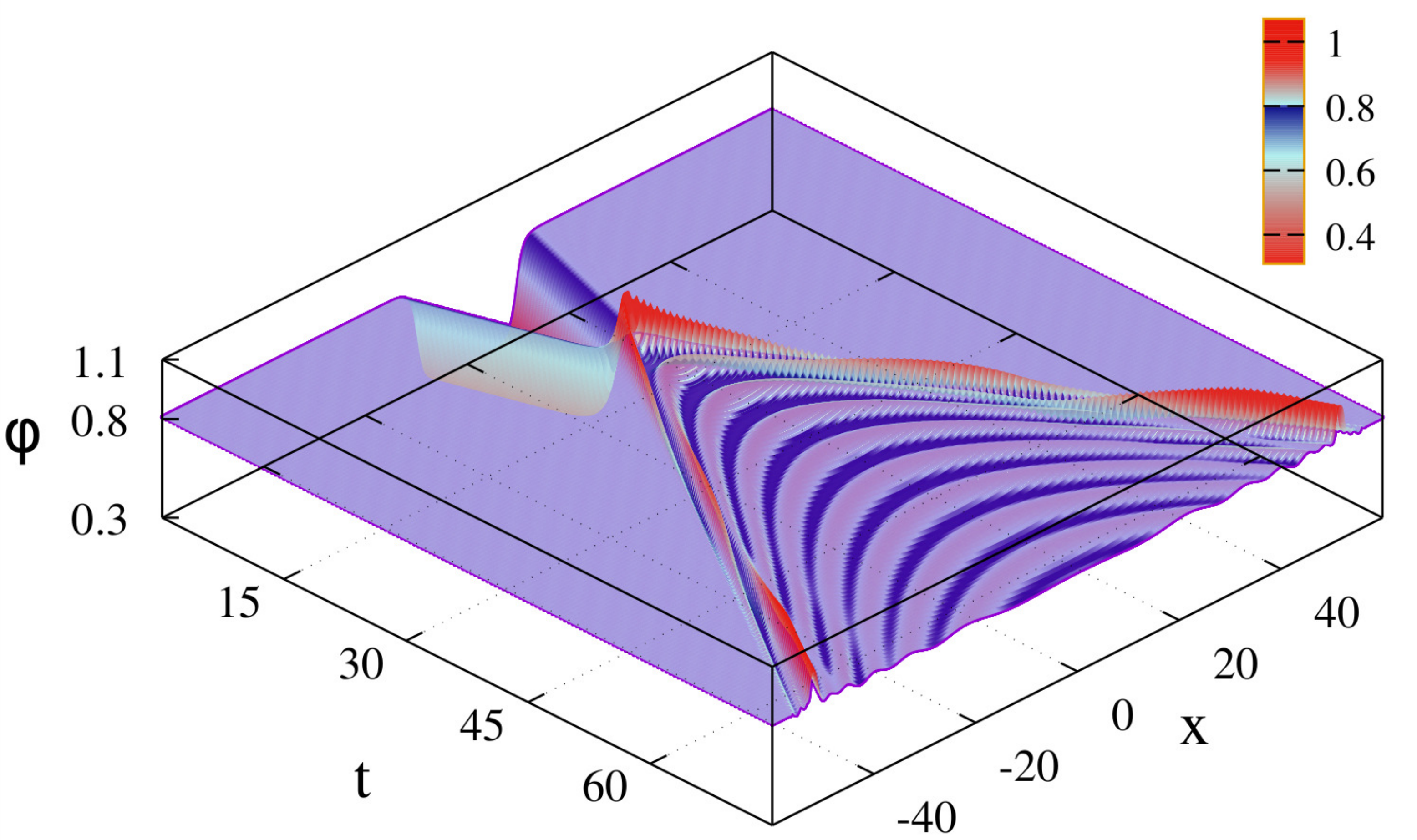}\label{fig:2kP08P03P08vi050vf092405}}
\subfigure[\:$v_i=0.9$, $v_f^{}=0.977$]{\includegraphics[width=0.45
 \textwidth]{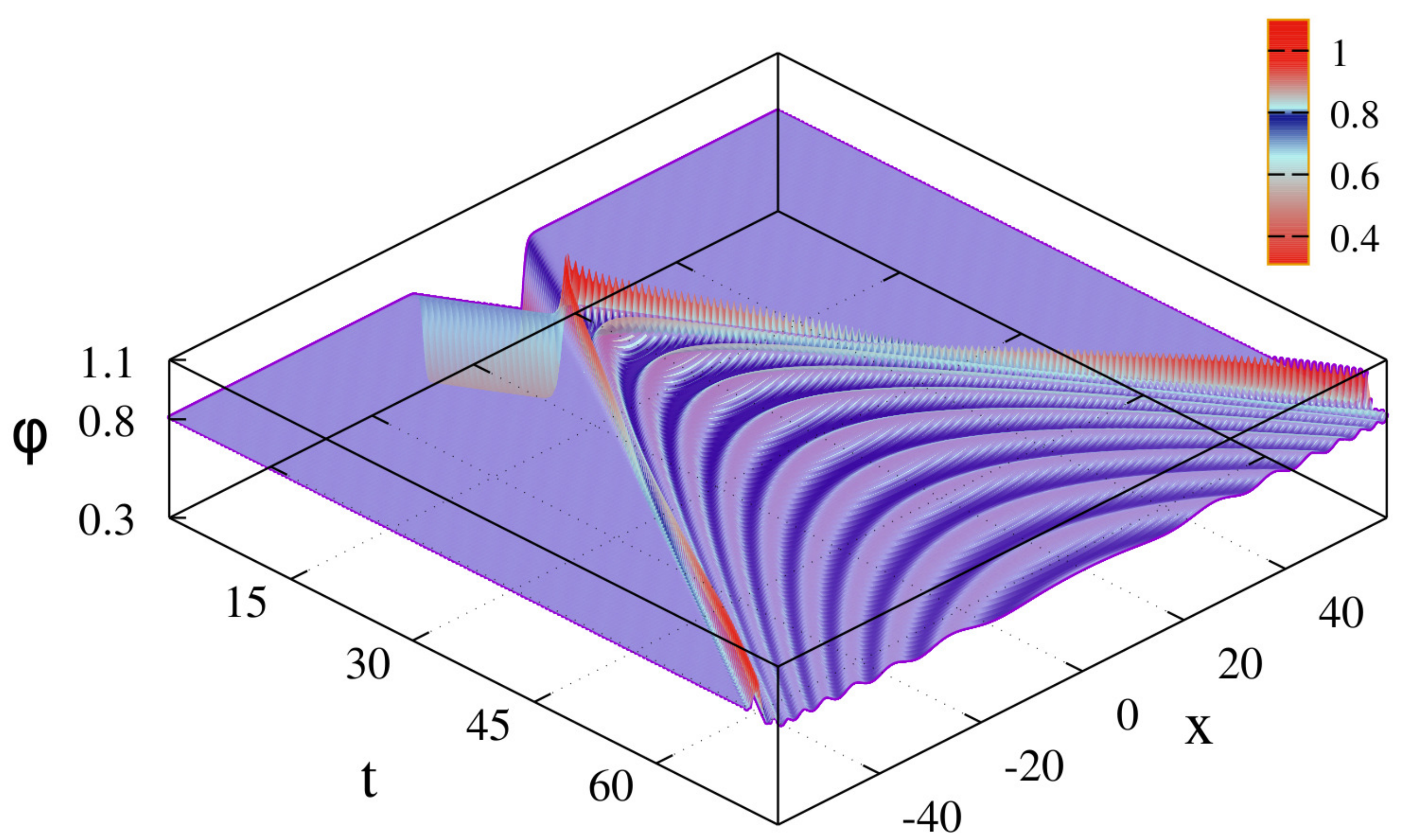}\label{fig:2kP08P03P08vi090vf097714}}
 \caption{Antikink-kink collision in the sector $(a,b)$.}
 \label{fig:2kFieldCaseII} 
 \end{figure}

An antikink-kink collision in the sector $(a,b)$ always gives rise to two high-speed oscillons and radiation, see Fig.~\ref{fig:2kFieldCaseII}.
Despite the fact that the production of a kink-antikink pair in the sector $(b,c)$ is energetically favorable, for some reason it does not occur.
Noteworthy that the ratio of the kink masses in the neighboring sectors $(a,b)$ and $(b,c)$ is large:
\begin{eqnarray}
    \frac{M_{\scriptsize\mbox{K}}^{(a,b)}}{M_{\scriptsize\mbox{K}}^{(b,c)}}=\frac{218}{93-25\sqrt{5}} \approx 5.876.
\end{eqnarray}

\subsection{Collision of kink and antikink in the sector  \texorpdfstring{$(b,c)$ }{pdfbookmark}}\label{sec:case II_}

Kink and antikink in this topological sector are the lightest among all kinks of the considered model. Again, kink-antikink and antikink-kink collisions will be considered separately.

\subsubsection{Kink-antikink collision}

\begin{figure}[t!] 
 \centering 
\subfigure[\:$v_i^{}=0.770$]{\includegraphics[width=0.45
 \textwidth]{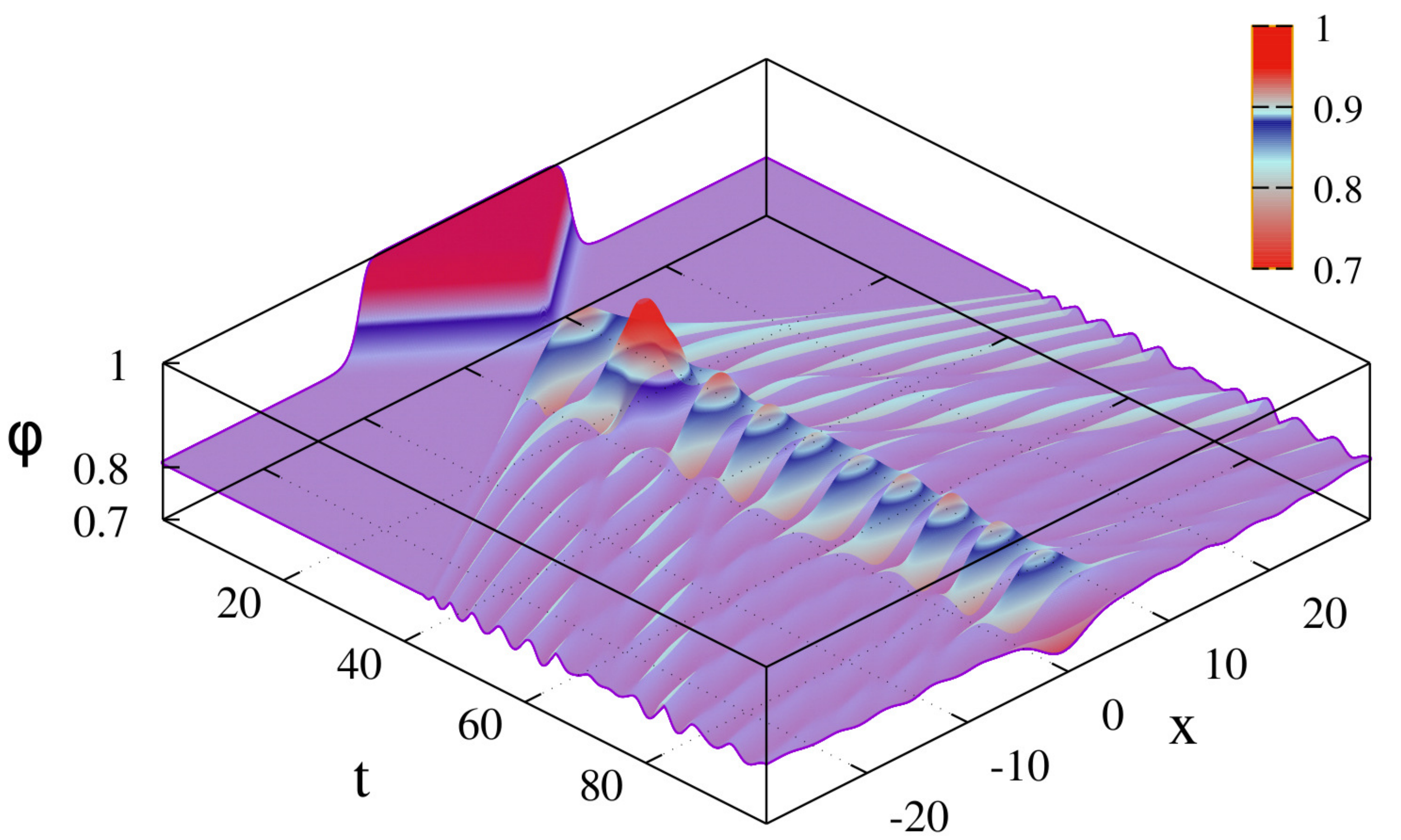}\label{fig:2kP08P10P08vi077vf00}}
\subfigure[\:$v_i^{}=0.800$, $v_f^{}=0.144$]{\includegraphics[width=0.45
 \textwidth]{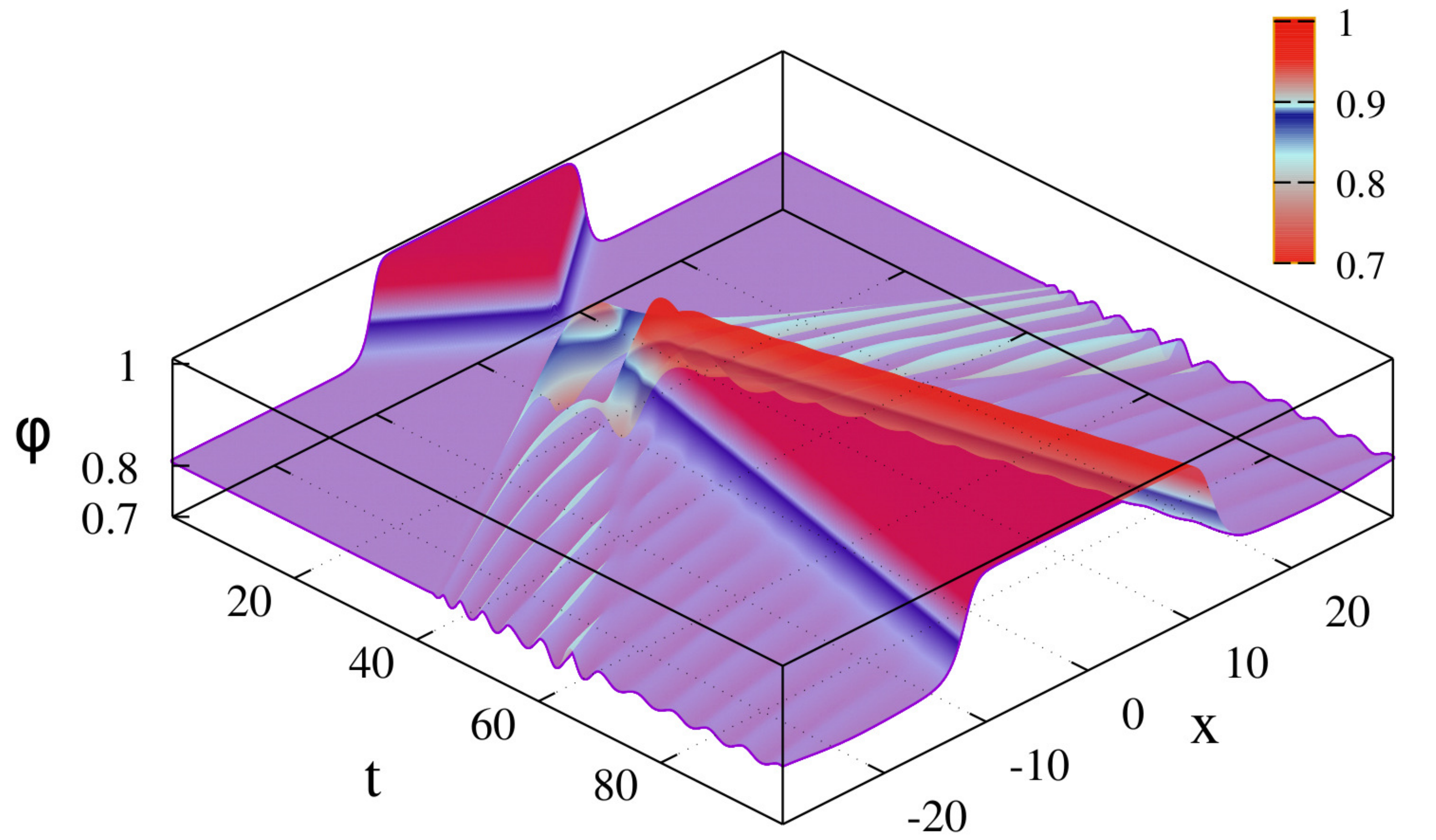}\label{fig:2kP08P10P08vi080vf01437}}
 \\
 \subfigure[\:$v_i^{}=0.900$, $v_f^{}=0.038$]{\includegraphics[width=0.45
 \textwidth]{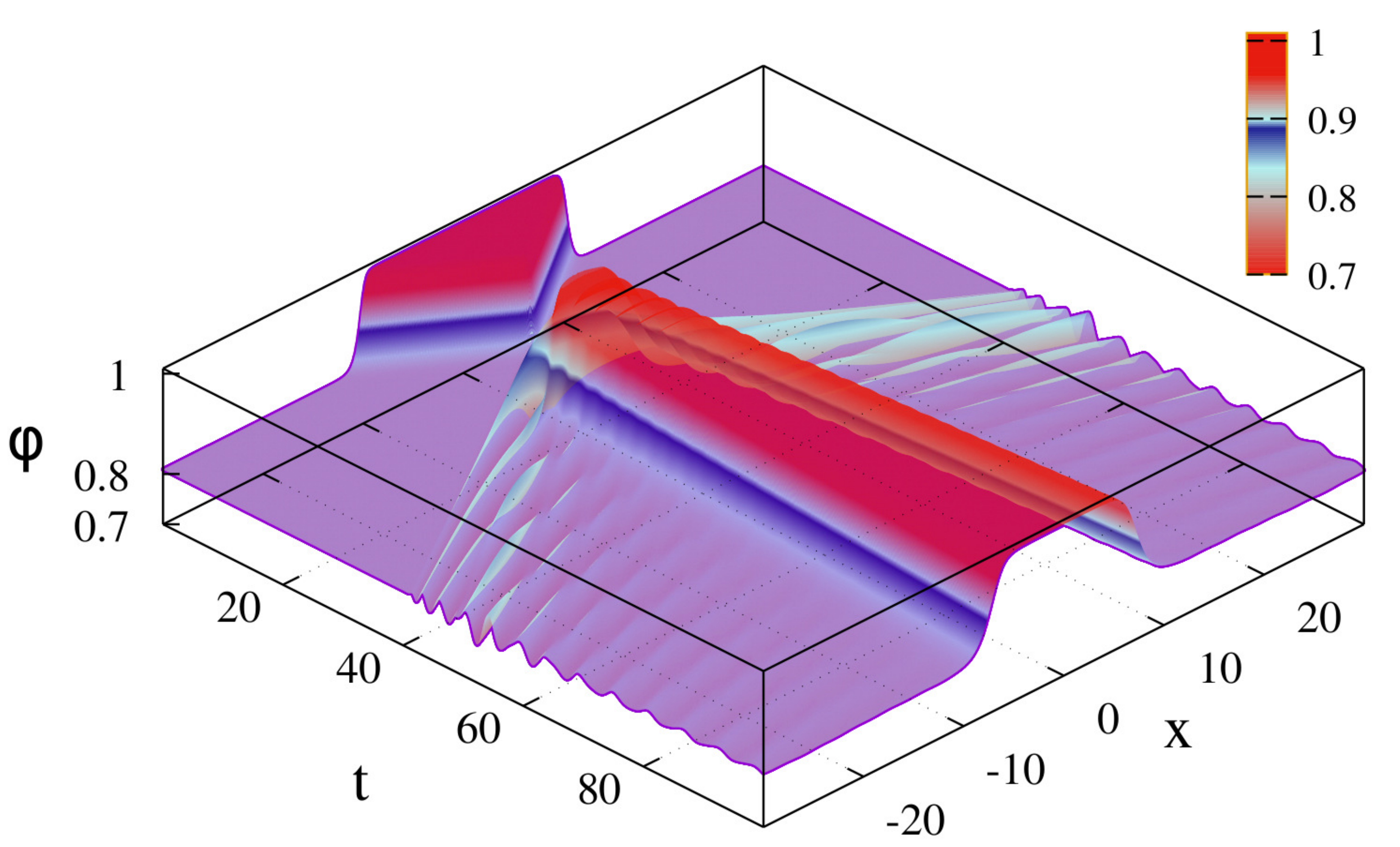}\label{fig:2kP08P10P08vi0900vf0038}}
 \caption{Kink-antikink collision in the sector $(b,c)$ at different initial velocities $v_i^{}$: (a) $v_i^{}<v_{cr}^{}=0.883$, a bion is formed; (b) two-bounce resonance escape at $v_i^{}=0.800$; (c) escape after one collision at $v_i^{}>v_{cr}^{}$.} 
 \label{fig:kinkantikinkFieldCaseIII} 
\end{figure}

\begin{figure}[t!]
    \centering
    \subfigure[]{\includegraphics[width=0.45 \textwidth, height=0.25 \textheight]{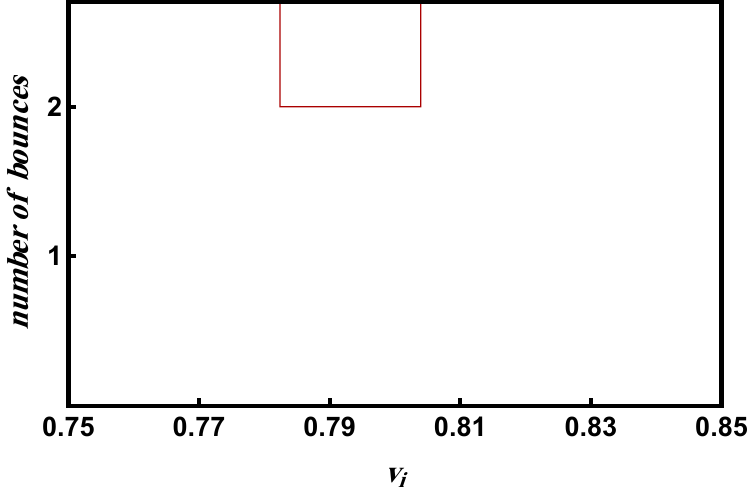}\label{fig:BouncesKAKcollisionUpsector}}
    \subfigure[]{\includegraphics[width=0.45\textwidth, height=0.25 \textheight]{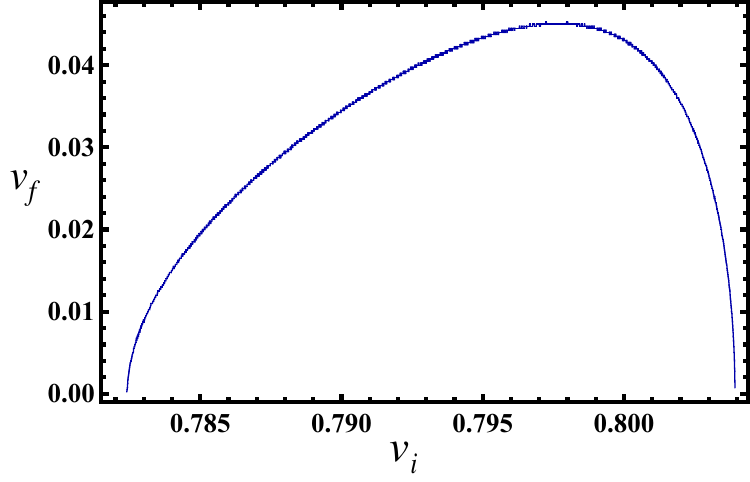}\label{fig:VfViKAKcollisionUpsector}}
    \caption{An example of the two-bounce escape window in the kink-antikink collision in the sector $(b,c)$. (a) Number of bounces as a function of the kink's initial velocity $v_i^{}$; (b) the final velocity $v_f^{}$ as a function of $v_i^{}$.
}
    \label{fig:KAKUPSectorNbScape}
\end{figure}

In this case, the critical velocity is $v_{cr}^{} = 0.883$: at $v_i^{}<v_{cr}^{}$ kink and antikink become trapped and form a bion at the collision point, see Fig.~\ref{fig:kinkantikinkFieldCaseIII}(a).
Moreover, near the critical velocity the so-called {\it escape windows} (or {\it bounce windows}) were observed. The escape window is a certain interval of the initial velocity from the range $v_i^{}<v_{cr}^{}$, within which resonant escape of colliding solitons occurs after two or more collisions. Fig.~\ref{fig:kinkantikinkFieldCaseIII}(b) illustrates the kink-antikink collision at the initial velocity from the {\it two-bounce window}. We emphasize that in the case of an escape window, the situation is fundamentally different from the escape after one impact, which occurs at $v_i^{} \geq v_{cr}^{}$, see Fig.~\ref{fig:kinkantikinkFieldCaseIII}(c). For a more detailed explanation of this phenomenon in other field-theoretic models see, e.g., \cite[Sec.~4]{Gani.JHEP.2015}, \cite[Sec.~4]{Gani.EPJC.2018.dsg}, \cite[Sec.~4]{Christov.CNSNS.2021}, \cite[Sec.~5]{Bazeia.EPJC.2018.sinh}, \cite[Sec.~3]{Belendryasova.CNSNS.2019}. In Fig.~\ref{fig:KAKUPSectorNbScape}
we give the dependence of the kink's final velocity on its initial velocity, thereby illustrating behavior in a two-bounce window.

\begin{figure}[t!] 
 \centering 
\subfigure[]{\includegraphics[width=0.45
 \textwidth,height=0.20 \textheight]{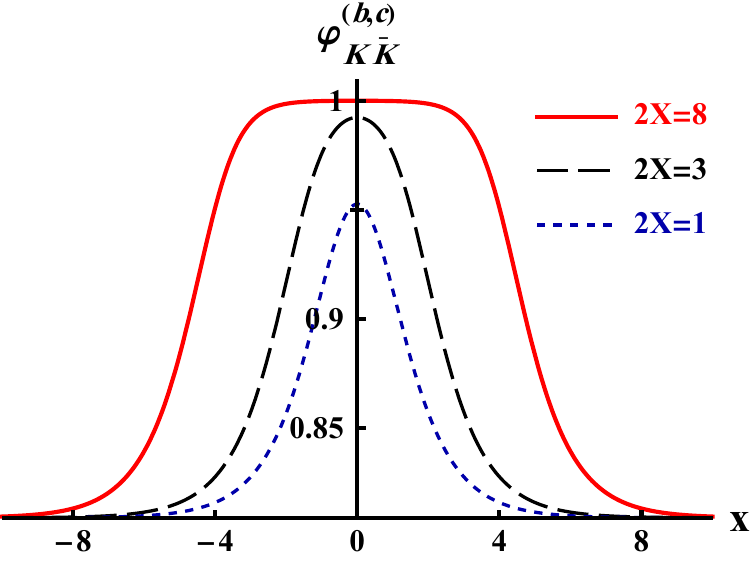}\label{fig:kakconfigurationIII}}
\subfigure[]{\includegraphics[width=0.45
 \textwidth,height=0.20 \textheight]{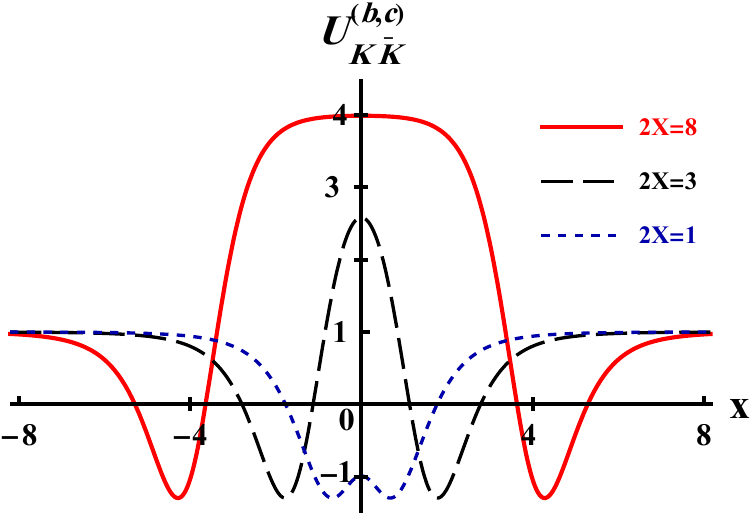}\label{fig:qmpconfigurationIII}}
 \caption{(a) Kink-antikink configurations in the sector $(b,c)$ for different separations $2X$. (b) The corresponding stability potentials.} 
 \label{fig:qmpkinkantikinkFieldCaseIII} 
\end{figure}

It is important to note that the kinks we are considering do not have vibrational modes that could act as energy accumulators. Nevertheless, resonant energy exchange takes place. Similar situation was observed in the $\varphi^6$ and $\varphi^8$ models \cite{Dorey.PRL.2011,Belendryasova.CNSNS.2019}. For the ``kink + antikink'' system in the sector $(b,c)$, the potential that determines the spectrum of small excitations in the linear approximation (also often called {\it stability potential}) is shown in Fig.~\ref{fig:qmpkinkantikinkFieldCaseIII}(b), together with the corresponding kink-antikink configurations, Fig.~\ref{fig:qmpkinkantikinkFieldCaseIII}(a).
From this figure it is seen that at small kink-antikink separation, the stability potential generated by the field configuration can look as a potential well. The time-independent Schr\"odinger equation with such a potential could have level(s) in the discrete spectrum, which could play role of vibrational mode(s) of the ``kink + antikink'' system as a whole. To confirm this hypothesis, on the one hand, a detailed study of the behavior of the field during the collision of solitons is required, and, on the other hand, an analysis of the discrete spectrum in the potential well (Fig.~\ref{fig:qmpkinkantikinkFieldCaseIII}(b)) depending on the distance between solitons should be done. The study of these questions, however, is beyond the scope of this paper.

Finally, it is noteworthy that at $v_i^{}\ge v_{cr}^{}$ the final velocities of the colliding kinks is much less than the initial velocities, see Fig.~\ref{fig:kinkantikinkFieldCaseIII}(c), and even at high velocities, we did not observe formation of an antikink-kink pair in the sector $(a,b)$ in the final state.

\subsubsection{Antikink-kink collision}

\begin{figure}[t!] 
 \centering 
 \subfigure[\:$v_i^{}=0.0700$, formation of bion]{\includegraphics[width=0.45
 \textwidth]{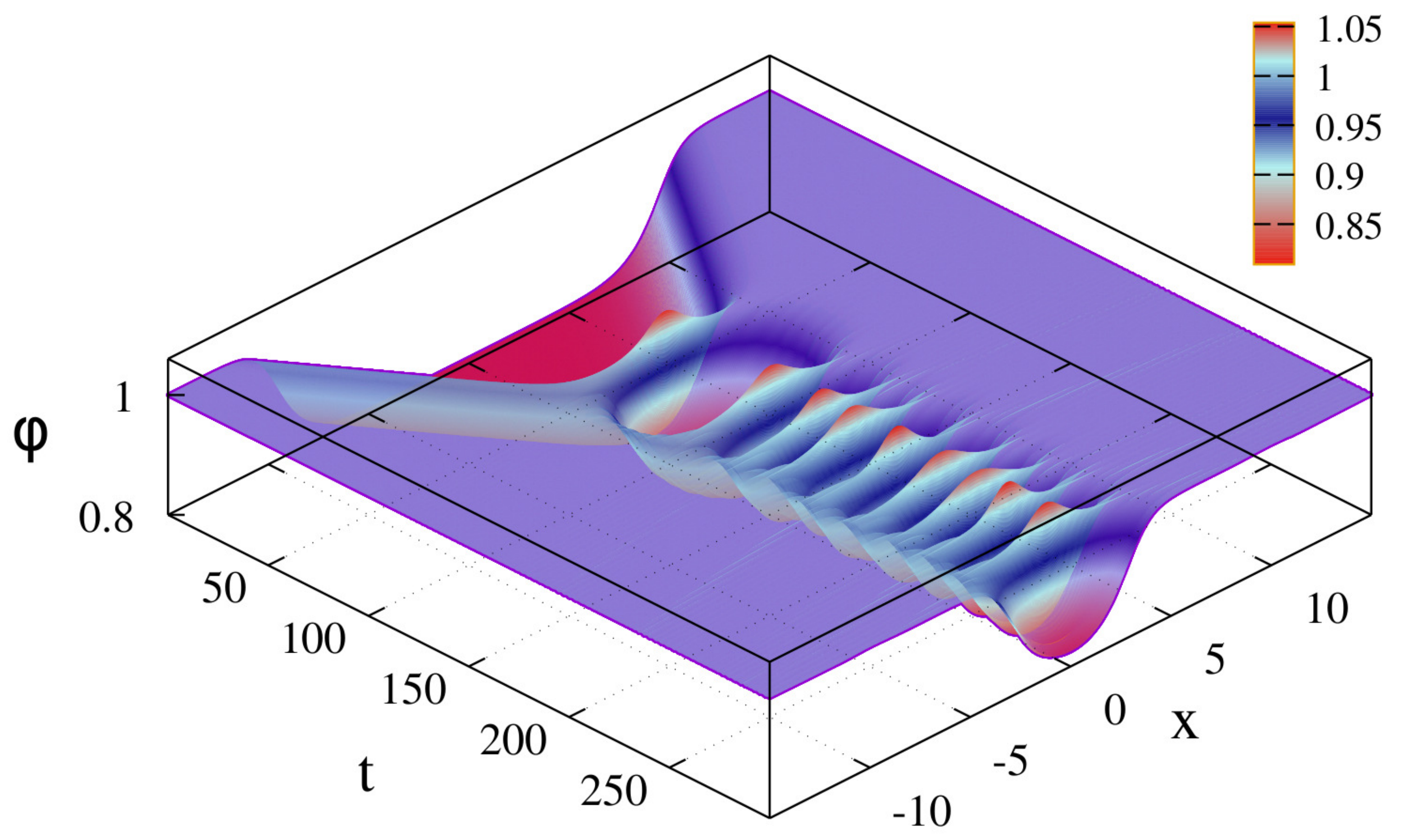} \label{fig:2kP10P08P10vi00700bn}} 
\subfigure[\:$v_i^{}=0.0894$, five-bounce escape window]{\includegraphics[width=0.45
 \textwidth]{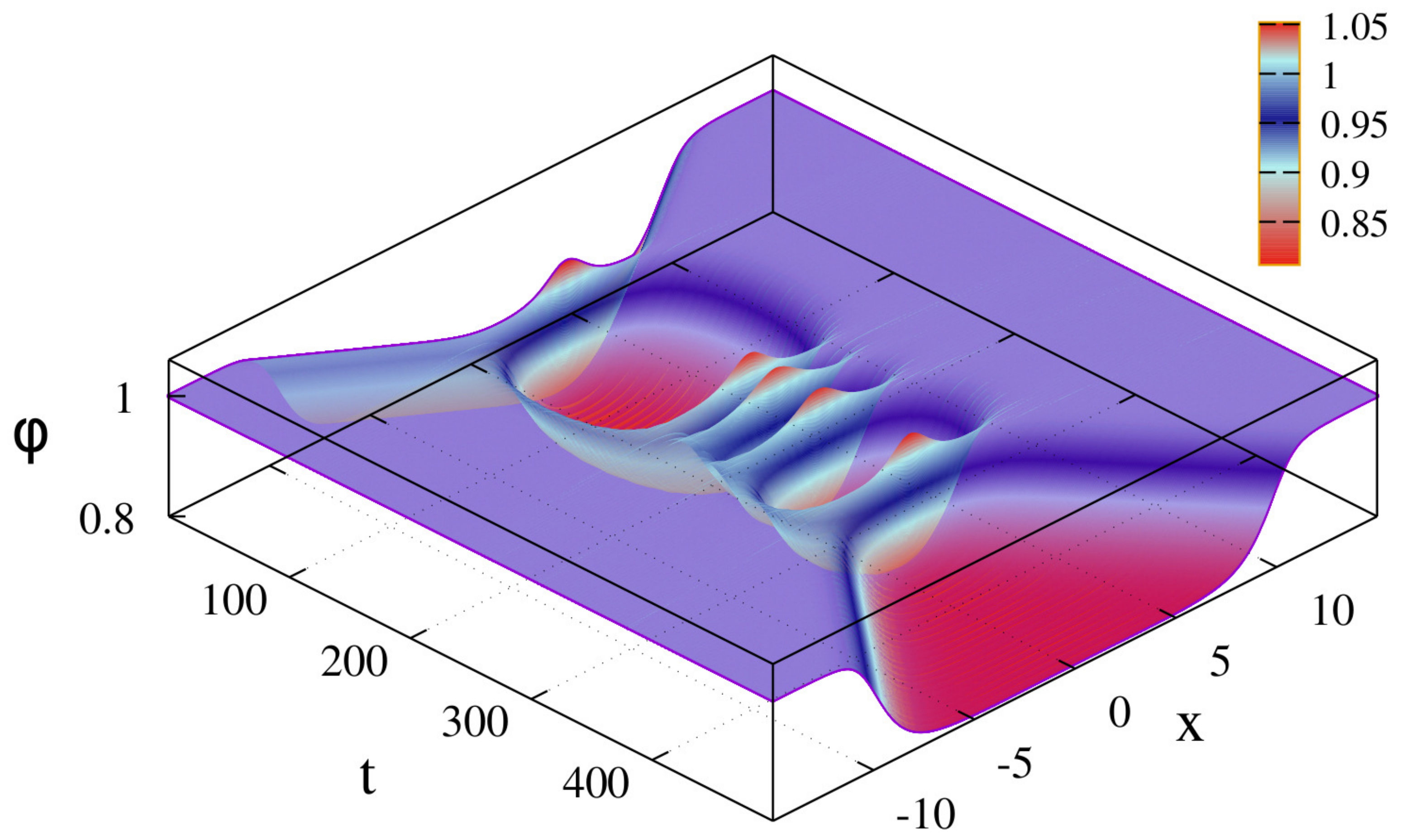}\label{fig:2kP10P08P10vi00894b5}}
 \\
\subfigure[\:$v_i^{}=0.0895$, two-bounce escape window]{\includegraphics[width=0.45
 \textwidth]{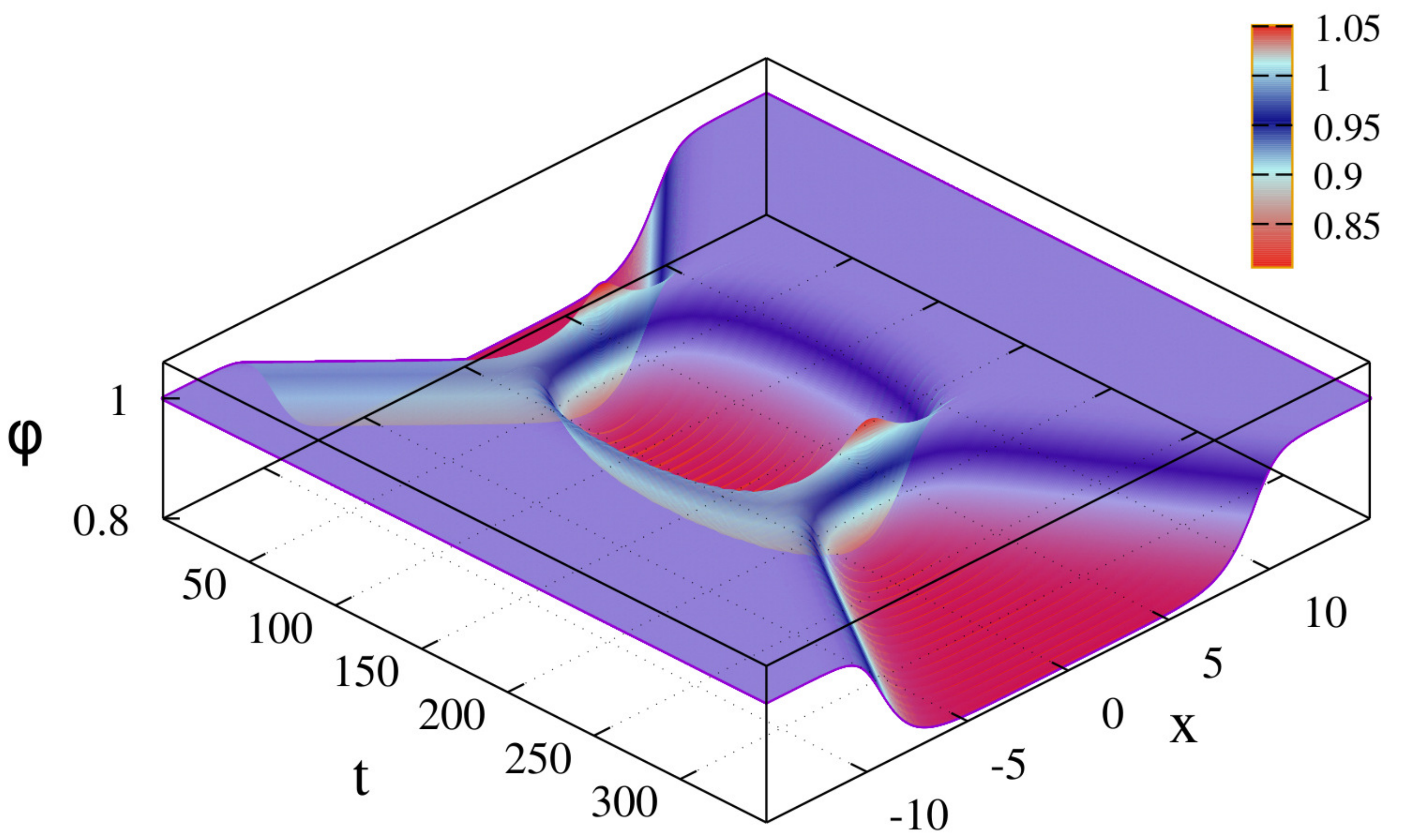}\label{fig:2kP10P08P10vi00895b2}}
\subfigure[\:$v_i^{}=0.0898$, three-bounce escape window]{\includegraphics[width=0.45
 \textwidth]{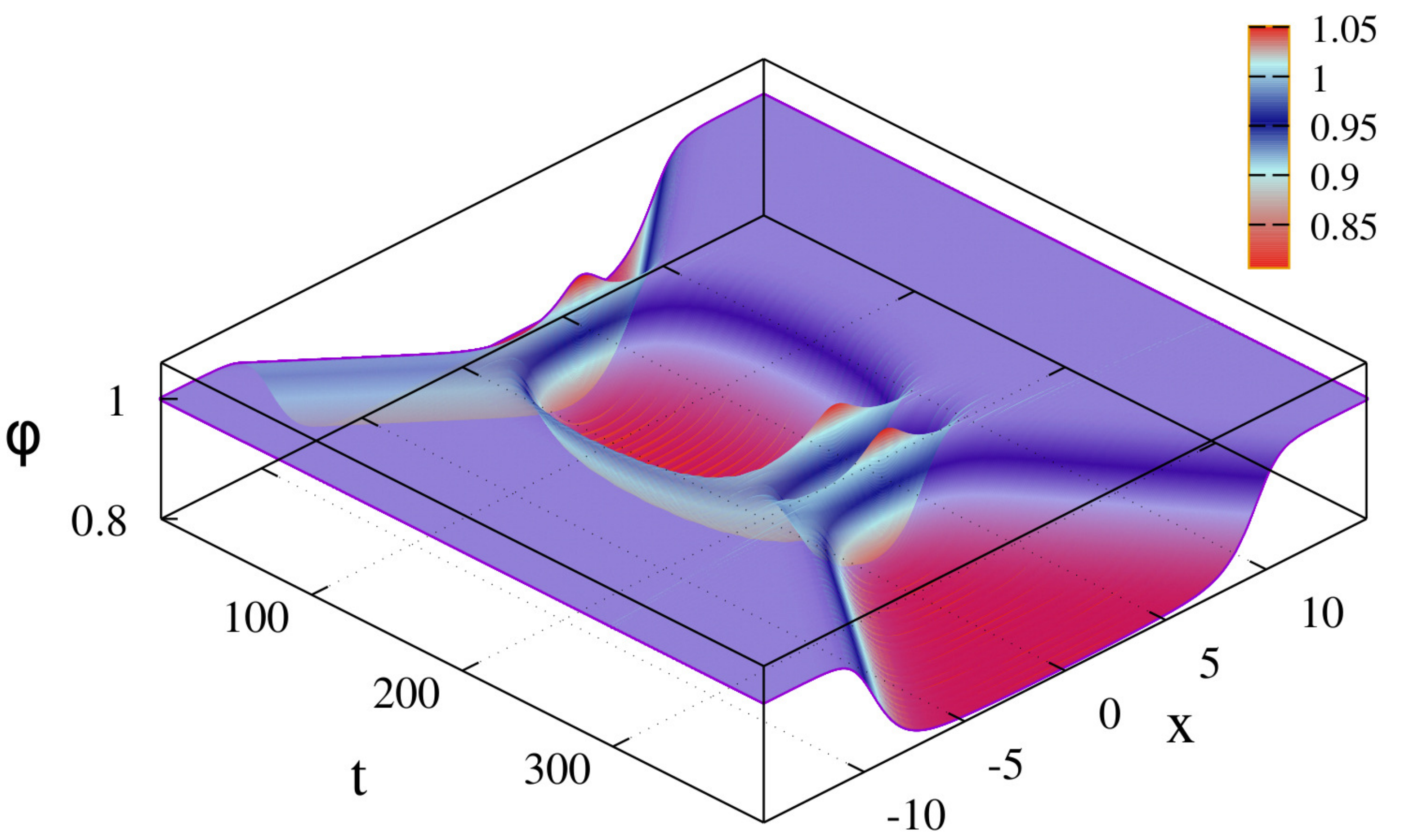}\label{fig:2kP10P08P10vi00898b3}}
\subfigure[\:$v_i^{}=0.0920$, escape after one collision]{\includegraphics[width=0.45
 \textwidth]{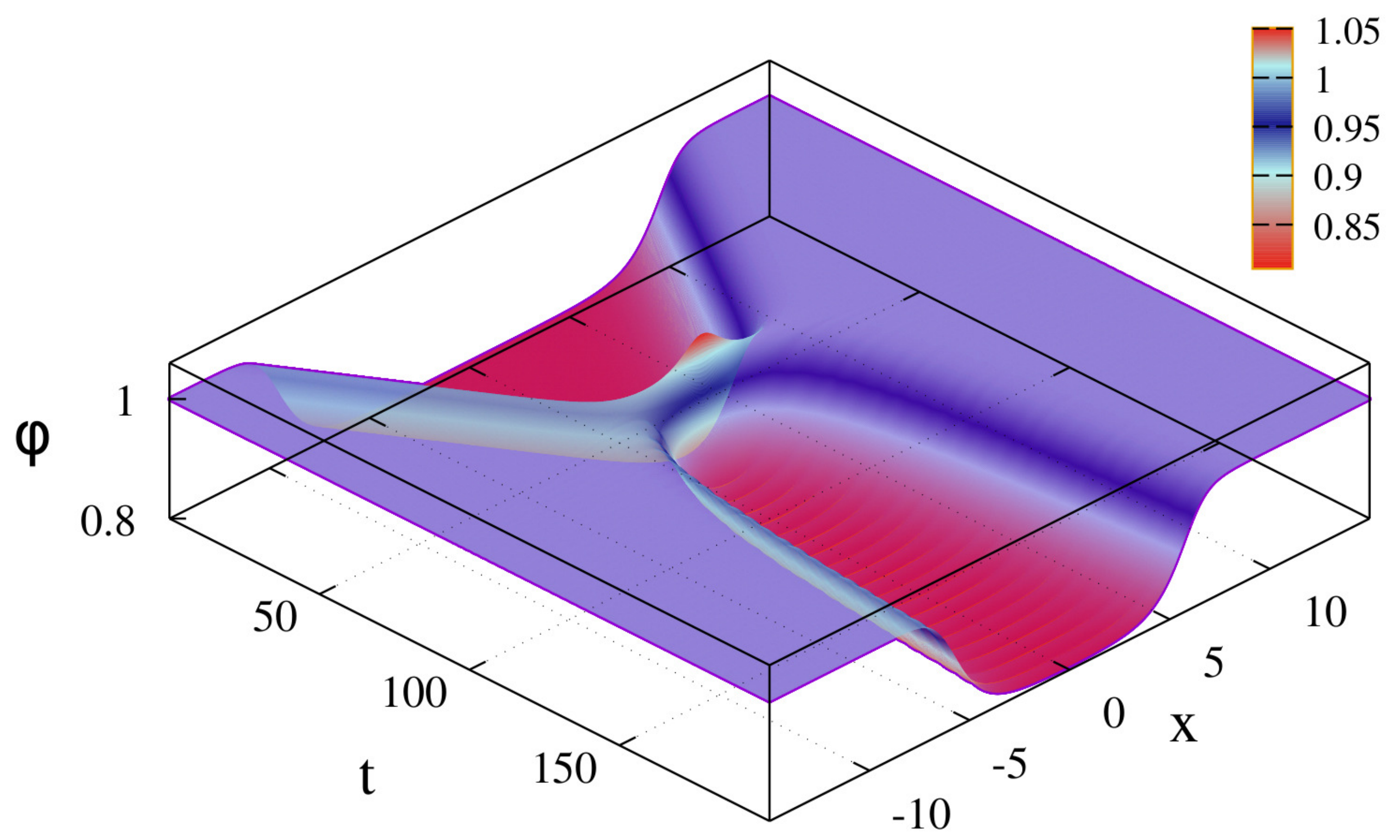}\label{fig:2kP10P08P10vi00920b1}}
 \caption{Antikink-kink collisions in the sector $(b,c)$.} 
 \label{fig:bouncessectorIII} 
\end{figure}

\begin{figure}[t!]
    \centering
    \subfigure[]{\includegraphics[width=0.45 \textwidth, height=0.25 \textheight]{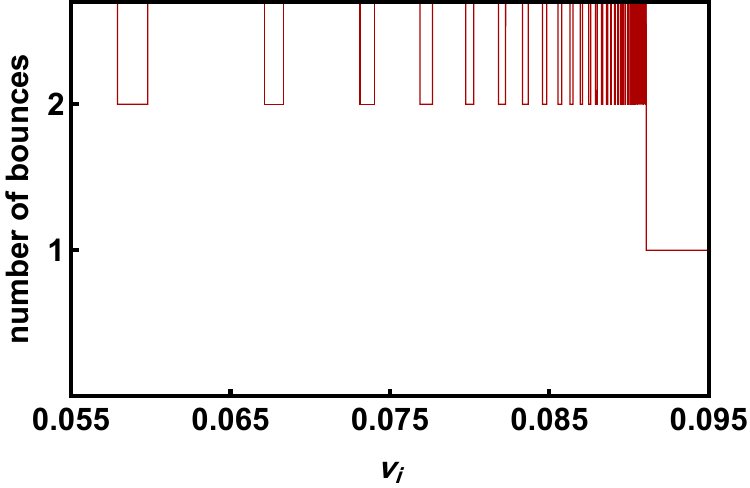}\label{fig:BouncesAKKcollisionUpsector}}
    \subfigure[]{\includegraphics[width=0.45\textwidth, height=0.25 \textheight]{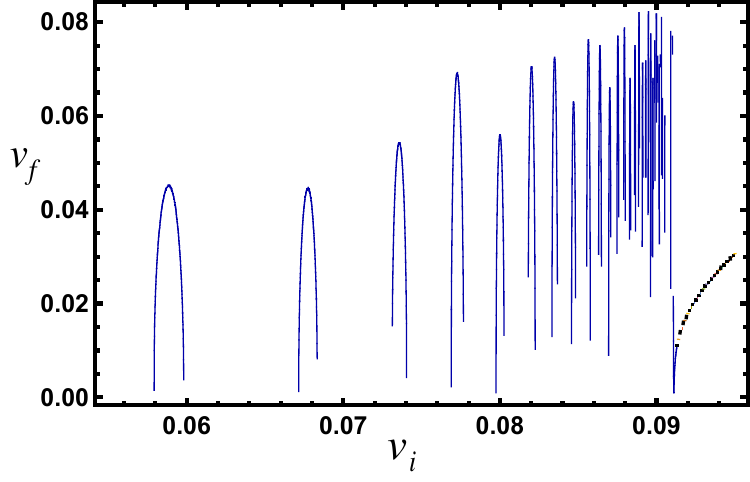}\label{fig:VfViAKKcollisionUpsector}}
    \caption{A series of the two-bounce escape windows in the antikink-kink collisions in the sector $(b,c)$. (a) Number of bounces as a function of the kink's initial velocity $v_i^{}$; (b) the final velocity $v_f^{}$ as a function of $v_i^{}$ (the two-bounce escape windows --- blue solid curves, and one-bounce escape at $v_i^{}>v_{cr}^{}$ --- black dotted curve).
    }
    \label{fig:AKKUPSectorNbScape}
\end{figure}

Antikink-kink collisions also demonstrate different regimes depending on the initial velocity: (i) at $v_i^{}<v_{cr}^{}=0.0913$, a bion is formed (Fig.~\ref{fig:bouncessectorIII}(a)), and escape windows are also present (Figs.~\ref{fig:bouncessectorIII}(b), \ref{fig:bouncessectorIII}(c), and \ref{fig:bouncessectorIII}(d));
(ii) at $v_i^{} \geq v_{cr}^{}$, the escape of kinks to spatial infinities is observed after only one collision (Fig.~\ref{fig:bouncessectorIII}(e)). The process in Fig.~\ref{fig:bouncessectorIII}(b) is a resonant escape of the solitons corresponding to the five-bounce escape window --- the antikink and kink collide exactly five times before escaping to spatial infinities. Figs.~\ref{fig:bouncessectorIII}(c) and \ref{fig:bouncessectorIII}(d) illustrate two- and three-bounce escape windows, respectively. In Fig.~\ref{fig:AKKUPSectorNbScape}
we give the dependence of the kink's final velocity on its initial velocity, thereby illustrating behavior in a series of the two-bounce windows.

It is natural to assume that the role of the energy accumulator in these processes is played by the vibrational mode(s) of the ``antikink + kink'' system as a whole, i.e.\ the mechanism discovered earlier in the $\varphi^6$ and $\varphi^8$ models \cite{Dorey.PRL.2011,Belendryasova.CNSNS.2019} works.

\section{Conclusion}
\label{sec:Conclusion}

We have studied some dynamic properties of topological solitons (kinks) of the (1+1)-dimensional $\varphi^{12}$ model. The considered model is remarkable in that it has five topological sectors --- one symmetric and four asymmetric.

First of all, we found the masses and asymptotics of all kinks, and obtained asymptotic estimates for the interaction force of a kink and an antikink separated by a large distance. As expected, in all cases the force decays exponentially with distance. At the same time, the decay rate differs depending on the asymptotic behavior of the kink field at large distances from the kink core. We have also experimentally measured the force of interaction between kink and antikink in various topological sectors. Comparison of experimental data with asymptotic estimates shows fairly good agreement, see Fig.~\ref{fig:Forces}(b).

Then, we have numerically studied the kink-antikink and antikink-kink scattering in different topological sectors.

\begin{itemize}

    \item In the collision of kink and antikink in the symmetric sector $(-a,a)$ we observed annihilation and production of a new antikink-kink pair in the sector $(-b,-a)$. This happens because the mass of the kink in sector $(-b,-a)$ is 1.5 times less than the mass of the kink in sector $(-a,a)$. This, in particular, leads to the fact that even taking into account the energy losses due to radiation, the velocities of the escaping kinks are higher than the velocities of the colliding kinks.

    \item In the asymmetric sector $(a,b)$ the kink-antikink and antikink-kink collisions occur differently.

    \begin{itemize}

        \item In the kink-antikink collisions, depending on the initial velocity, three different regimes were found: (i) at $v_i^{} < v_{cr}^{}=0.442$, we observed the formation of a bion at the collision point; (ii) at $v_{cr}^{} \leq v_i^{} < v_{cs}^{}=0.782$, the kinks did not change their topological sector; (iii) at $v_i^{} \geq v_{cs}^{}$, the colliding solitons annihilated and an antikink-kink pair in the sector $(-a,a)$ was formed instead.

        \item In the antikink-kink collisions we observed only annihilation into a pair of high-speed oscillons and radiation.
        
    \end{itemize}

    \item In the asymmetric sector $(b,c)$ the kink-antikink and antikink-kink collisions also occur differently.

    \begin{itemize}

        \item In the kink-antikink collisions, first, a critical velocity $v_{cr}^{} = 0.883$ was found, below which kinks are captured and form a decaying bound state, and above which they scatter to spatial infinities after one collision. Second, escape windows --- intervals of initial velocity within which kinks escape to spatial infinity after two or more collisions --- were found below the critical velocity. Escape windows are a typical resonance phenomenon resulting from the exchange of energy between the translational and vibrational modes of kink. Interestingly, in this case, there are no vibrational modes in the excitation spectrum of the kink. As we already mentioned above, cases are known when not the vibrational modes of a separate kink and antikink act as an energy accumulator, but their collective modes, which arise only when the kink and antikink are close together during a collision.

        \item In the antikink-kink collisions, a critical velocity $v_{cr}^{}=0.0913$ was also found. At $v_i^{}<v_{cr}^{}$, the kinks are captured, forming a bion, or a resonant escape occurs, corresponding to escape windows. At $v_i^{} \geq v_{cr}^{}$, the kinks escape to spatial infinities after one collision.

    \end{itemize}

\end{itemize}

Finally, in our opinion, the study carried out by us and presented in this paper may have interesting continuations.

\begin{enumerate}

    \item In the collisions of the $\varphi^{12}$ kinks, escape windows were observed. Their appearance means that there is a resonant energy exchange between the translational mode (kinetic energy) of kinks and a certain energy accumulator. In many well-known cases, the vibrational mode of the kink played the role of such an accumulator. However, all kinks in the considered model do not have vibrational modes. On the other hand, the stability potentials of asymmetric kinks are asymmetric, which can lead to the appearance of vibrational modes of the ``kink+antikink'' or ``antikink+kink'' systems as a whole. This possibility may become the subject of a separate study.

    \item In the antikink-kink collisions in the sector $(a,b)$, lighter kink-antikink pair in the sector $(b,c)$ is not produced. The reason for this is not entirely clear, and its elucidation may be the subject of future research.

\end{enumerate}

\section*{Acknowledgments}
AMM would like to thank Islamic Azad University Quchan branch for the grant. The research was carried out within the state assignment of Ministry of Science and Higher Education of the Russian Federation, project  No.\ FSWU-2023-0031. A.~Ghaani thanks the Ferdowsi University of Mashhad for Post Doctoral support under the scientific authority program.

\end{document}